\documentclass[a4paper,11pt]{article}
\pdfoutput=1 
\usepackage[symbol]{footmisc} 
\usepackage{jheppub} 

\usepackage{color, colortbl}
\usepackage[T1]{fontenc}
\usepackage[utf8]{inputenc}
\usepackage[T1]{fontenc}
\usepackage{epsfig,latexsym}
\usepackage{amsmath}
\usepackage[dvipsnames]{xcolor}
\usepackage{adjustbox}
\usepackage{tensor}
\usepackage{graphicx}
\usepackage{verbatim}
\usepackage{mathrsfs}
\usepackage{amssymb}
\usepackage{multirow}
\usepackage{epsfig}
\usepackage{color,colordvi}
\usepackage{appendix}
\usepackage{slashed}
\usepackage{array}
\usepackage{cancel}
\usepackage{float}
\usepackage[font=footnotesize, justification=centering]{caption}
\usepackage{epsf}
\usepackage{amsmath}
\usepackage{physics}
\usepackage{MnSymbol}
\usepackage{rotating}
\usepackage{multirow}
\usepackage[normalem]{ulem}
\usepackage{tcolorbox}
\usepackage{bbold}
\usepackage{nicefrac}
\DeclareMathAlphabet{\mathpzc}{OT1}{pzc}{m}{it}

 \csname
@addtoreset\endcsname{equation}{section}

\newcolumntype{x}[1]{>{\centering\arraybackslash\hspace{0pt}}p{#1}}

\newcommand{\beq}{\begin{equation}}
\newcommand{\eeq}{\end{equation}}
\renewcommand{\[}{\left[}
\renewcommand{\]}{\right]}
\renewcommand{\(}{\left(}
\renewcommand{\)}{\right)}

\newcommand{\be}{\begin{eqnarray}}
\newcommand{\ee}{\end{eqnarray}}
\newcommand{\bea}{\begin{eqnarray}}
\newcommand{\eea}{\end{eqnarray}}
\newcommand{\bi}{\begin{itemize}}
\newcommand{\ei}{\end{itemize}}
\newcommand{\ben}{\begin{enumerate}}
\newcommand{\een}{\end{enumerate}}
\def\bes{\begin{equation*}}
\def\ees{\end{equation*}}
\def\bead{\begin{aligned}}
\def\eead{\end{aligned}}
\def\bmat{\left(\begin{matrix}}
\def\emat{\end{matrix}\right)}

\newcommand{\N}{\mathcal{N}}
\newcommand{\F}{\mathcal{F}}
\newcommand{\A}{\mathcal{A}}

\newcommand{\G}{\mathcal{G}}
\renewcommand{\S}{\mathcal{S}}

\newcommand{\bpsi}{\bar{\psi}}

\newcommand{\bepsilon}{\bar{\epsilon}}

\newcommand{\bchi}{\bar{\chi}}
\newcommand{\bp}{\boldsymbol{p}}

\renewcommand{\[}{\left[}
\renewcommand{\]}{\right]}
\renewcommand{\(}{\left(}
\renewcommand{\)}{\right)}

\newcommand{\mpsi}{\lambda}

\title{ \center A unique coupling of \\
the massive spin-2 field to supergravity}

\author{Guillaume Bossard,}
\author{Gabriele Casagrande,}
\author{Emilian Dudas}
\author{and Adrien Loty}
\affiliation{Centre de Physique Th{\'e}orique, {\'E}cole Polytechnique, CNRS and IP Paris, 91128 Palaiseau Cedex, France }

\emailAdd{guillaume.bossard@polytechnique.edu, gabriele.casagrande@polytechnique.edu, emilian.dudas@polytechnique.edu, adrien.loty@polytechnique.edu}

\abstract{ We show that the coupling of a massive spin-2 field to undeformed $\mathcal{N}=1$ supergravity in four dimensions is unique, leading to a specific non-minimal coupling to the Riemann tensor. The massive spin-2 coupling reproduces the one of an oscillator mode in open string theory, while the massive spin-1 coupling includes a higher-derivative term that is expected to violate causality in the background of a gravitational shock wave. We argue that the resolution of causality and the unitarity bound in the Regge limit require the introduction of infinitely many higher-spin fields similar to the Regge trajectories in string theory, therefore providing an argument in favour of the string lamppost principle with minimal supersymmetry in four dimensions. To obtain this result, we construct the general stress-energy tensor multiplet for the massive spin-2 multiplet with $\mathcal{N}=1$ supersymmetry. }

\begin{document}
\begin{flushright}
CPHT-RR002-012025
\end{flushright}

\maketitle

\flushbottom


\section{Introduction}\label{sec:intro}
String theories generally include massive spin-2 particles, with no scale separation between the lightest massive spin-2 particle and the heavier particles of spin $s\ge 2$. These massive spin-2 particles are either Kaluza--Klein modes of the graviton, or string oscillator modes that come together with  a Regge trajectory of higher-spin fields. A natural question within the swampland program \cite{Vafa:2005ui,Palti:2019pca} is to wonder if this is a peculiarity of string theory or if this is a requirement any consistent theory of quantum gravity should satisfy. It has been suggested, at least assuming supersymmetry, that all such quantum theories of gravity may be realised as string theories \cite{Adams:2010zy,Kim:2019ths,Kim:2019vuc,Bedroya:2021fbu,Montero:2020icj}. If one follows this string lamppost paradigm, one should be able to prove that a massive spin-2 particle cannot appear without an infinite tower of massive states. In this trend of ideas, it has been argued that there is no consistent low energy effective theory below the scale $\Lambda$ including finitely many particles with only one spin-2 particle of mass $m\ll \Lambda$ and no higher-spin particle with mass $m_s < \Lambda$ and spin $s\ge 2$ \cite{Klaewer:2018yxi,Kundu:2023cof}. 

The main argument comes from the study of the elastic two-to-two scattering amplitude between two massive particles of spin-2. Although no actual proof exists for ungapped theories, unitarity bounds on the local growth of the scattering amplitudes are still expected to hold in the Regge limit \cite{Camanho:2014apa,Haring:2022cyf}. In particular, it has been proposed that the tree-level scattering amplitude of a single massive spin-2 field should not grow faster than $s^2$ in the Regge limit \cite{Chowdhury:2019kaq}, or more precisely for all $\epsilon>0$
\beq
    \lim_{s\to\infty}\frac{\A(s,t)}{s^{2+\epsilon}}=0\; . 
\label{eqn:crg}
\eeq
This conjecture was originally based on the bound on chaos discussed in  \cite{Maldacena:2015waa}, and was strengthened using the analyticity of the S-matrix for $D>4$ in \cite{Haring:2022cyf}. This bound is in particular saturated by the tree-level Kaluza--Klein amplitude that grows as $s^2$, when including the interaction with the higher Kaluza--Klein modes of mass $m_2 = 2 m$ \cite{SekharChivukula:2019yul,Chivukula:2020hvi,Bonifacio:2019ioc,Quirant:2024ian}. Building on some earlier work \cite{Bonifacio:2019mgk}, it has been shown in \cite{Kundu:2023cof} that there is no two-to-two tree-level scattering amplitude for a massive spin-2 particle satisfying \eqref{eqn:crg}, assuming there is no other particle of spin $s\ge 2$ and that the spin-2 particle interacts with Einstein gravity.

Defining the cutoff scale as the mass scale of the higher-derivative expansion in the low energy effective action, the cutoff scale is related to the higher-energy limit of the two-to-two scattering amplitude at fixed scattering angle $ \A(s,\theta )\sim s^n$ by 
\begin{equation}
    \Lambda_n= \(m^{n-1}M_{\scalebox{0.6}{P}}\)^{\frac1n}\; , 
\label{eqn:cutoff}
\end{equation}
where $m$ is the spin-2 mass and $M_{\scalebox{0.6}{P}}$ is the Planck mass \cite{Arkani-Hamed:2002bjr}. The higher the power $n$, the sooner the theory will enter into the strong-coupling regime, violating the effective field theory approximation. The elastic scattering amplitude in the theory minimally coupled to gravity grows as $s^5$, which gives the power-counting cutoff scale $\Lambda_5=\(m^4M_{\scalebox{0.6}{P}}\)^{\frac15}$ \cite{Arkani-Hamed:2002bjr}. Such rapid growth is due to the longitudinal polarisations of the massive spin-2 field, which dominate the amplitude at high-energy, as expressed in general terms by the equivalence theorem for massive gauge fields of any spin. Although it is raised by introducing further interactions, the cutoff cannot exceed $\Lambda_3$.
The impossibility of overcoming the $\Lambda_3$ scale has been proven for nonlinear deformations of the Fierz–Pauli theory \cite{Arkani-Hamed:2002bjr, Arkani-Hamed:2003roe, Schwartz:2003vj, Bonifacio:2018vzv}, as well as for the addition of couplings -- including higher-derivative ones -- with gravity \cite{Bonifacio:2018aon} and with lower-spin fields \cite{Bonifacio:2019mgk, Kundu:2023cof}. In particular, the theories that saturate the $\Lambda_3$ threshold are the de~Rham--Gabadadze--Tolley (dRGT) theory of massive gravity \cite{deRham:2010ik, deRham:2010kj} and the Hassan--Rosen bigravity theory \cite{Hassan:2011zd}, which are the known, ghost-free nonlinear theories of massive spin-2 fields, as proven in \cite{Bonifacio:2018vzv} and \cite{Bonifacio:2018aon} respectively.

However, the true cutoff scale, defined as the mass of the lightest particle that has been integrated out in the effective theory, does not necessarily saturate the upper bound cutoff scale defined by the higher-energy growth of the amplitude. The string lamppost principle mentioned above would predict instead that there is no good effective theory for a single massive spin-2 particle, or in other words that the true cutoff scale is of the order of the mass of the spin-2 particle, \textit{i.e.} $\Lambda_\infty$. For the dRGT theory of massive gravity, unitarity, causality and crossing symmetry imply that the cutoff is parametrically of the same order as the mass of the spin-2 particle \cite{Bellazzini:2023nqj}. The analysis of tree-level amplitudes in \cite{Kundu:2023cof} suggest that this is the case independently of the self-interactions, provided that the spin-2 field interacts with massless gravity. In this paper we wish to further argue that there is indeed no consistent effective theory for a single massive spin-2 particle coupled to (massless) $\mathcal{N}=1$ supergravity in four dimensions.

\vskip 4mm

In presence of rigid $\N=1$ supersymmetry, the massive spin-2 field is embedded in a supersymmetric multiplet together with a massive spin-1 and two massive Majorana spin-$\frac32$ fields. The free theory was constructed in \cite{Buchbinder:2002gh,Zinoviev:2002xn,Gates:2013tka} and represents the supersymmetric extension of the original Fierz--Pauli theory \cite{Fierz:1939ix}. Possible self-interactions, as well as extensions to different spacetime dimensions and to higher number of supercharges, were studied in \cite{DelMonte:2016czb, Zinoviev:2018juc, Engelbrecht:2022aao}. The simplest way to couple a massive spin-2 multiplet to gravity is to consider supergravity in higher dimension and compactify the theory in a way that only preserves $\mathcal{N}=1$ supersymmetry. The minimal example is defined by the Kaluza--Klein reduction of pure 5D supergravity on the orbifold $S^1/\mathbb{Z}_2$, where $\mathbb{Z}_2$ acts both as the reflection on the circle and as an $R$-symmetry to be a symmetry of supergravity in five dimensions. In this way one obtains $D=4$ supergravity coupled to a massless chiral multiplet and a tower of (real) massive spin-2 multiplets (see Appendix \ref{app:kk} for details). However, there is no consistent truncation of this theory to the first massive multiplet and there is no well defined effective theory with a separation of scale between the first Kaluza--Klein mode and the others. From the two-to-two amplitude one finds that the interaction with the next Kaluza--Klein mode is necessary to ensure the expected unitarity bound \eqref{eqn:crg} in the Regge limit, as one can see from the analysis of \cite{SekharChivukula:2019yul,Chivukula:2020hvi} in Einstein gravity in five dimensions. Moreover, the absence of propagating ghosts in the theory relies on the non-linear local supersymmetry gauge invariance coming from five-dimensional supergravity, which is broken in the truncation to the first Kaluza--Klein mode. The massive multiplet coming from the Kaluza--Klein reduction is indeed naturally defined in the St\"uckelberg formulation, together with additional pure gauge fields for the gauge invariance induced by diffeomorphism and local supersymmetry in five dimensions. The consistency of the theory therefore relies on five-dimensional gauge invariance. As we shall explain in Section  \ref{sec:stuckelberg}, this implies that the $\mathcal{N}=1$ supersymmetry algebra is deformed in the unitary gauge for the massive fields. Here we shall focus instead on massive spin-2 multiplets that can be coupled to supergravity in the unitary gauge without deforming the algebra. 

Our main result is that the coupling of a massive spin-2 multiplet in the unitary gauge to undeformed (off-shell)  supergravity is uniquely fixed at leading order and necessarily includes a higher-derivative coupling between the massive spin-1 field and the Riemann tensor of the type 
\be - \frac3{16m^2} R_{\mu\sigma\nu\rho} F^{\mu\sigma}F^{\nu\rho}\; . \label{RFF}\ee
According to \cite{Camanho:2014apa}, one expects such a coupling to lead to violations of causality that can only be resolved through the introduction of an infinite Regge trajectory of massive higher-spin fields. This result therefore suggests that the higher-spin Regge trajectory present in string theory can be predicted from the low energy effective action itself.\footnote{Meaning the infinity of higher-spin fields, not their specific spectrum.} There are in this way two kinds of massive spin-2 multiplets  in supergravity, the `Kaluza--Klein mode' and the `string oscillator'. We claim that unitarity can only be resolved  by the Kaluza--Klein tower in the former and by Regge trajectories in the latter.  

\vskip 4mm

At leading order in the Planck mass $M_{\scalebox{0.6}{P}}$, the coupling to $D=4$, (undeformed) $\N=1$ supergravity is given by the current of (rigid) supersymmetry and the stress-energy tensor. These currents are part of the so-called supercurrent superfield \cite{Ferrara:1974pz,Sohnius:1981tp,Komargodski:2010rb}. As discussed in the seminal paper \cite{Ferrara:1974pz} of Ferrara and Zumino, this multiplet of currents is a linear superfield that satisfies an additional conservation-like equation, which holds on-shell and defines its components. Together with the supersymmetry current and the stress-energy tensor -- which are always present -- the Ferrara--Zumino multiplet involves a pseudo-vector and a complex scalar field. A different supercurrent superfield can instead be defined for theories with an $R$-symmetry. This so-called $R$-multiplet satisfies a different characteristic equation, it contains the $R$-symmetry current as bottom component and includes an antisymmetric two-form field \cite{Sohnius:1981tp}. In the presence of full superconformal symmetry, the Ferrara--Zumino complex scalar, the divergence of the pseudo-vector and the $R$-multiplet two-form vanish, and the two supercurrent multiplets are identified as the multiplet of currents of the superconformal symmetry. Finally, there are instances in which neither the Ferrara--Zumino multiplet \cite{Ferrara:1974pz} nor the $R$-multiplet \cite{Sohnius:1981tp} can be consistently defined: in this cases one has to resort to a more general definition of the supercurrent superfield, interpolating between the two former ones \cite{Komargodski:2010rb}. Moving from rigid to local supersymmetry, the current of supersymmetry and the stress-energy tensor naturally couple to the supergravity fields, \textit{i.e.} the gravitino and the metric, as they are the currents of the gauge symmetry of supergravity. The current multiplets are dual to the supergravity multiplets and were used to derive the first off-shell formulations of supergravity \cite{Ferrara:1978em, Stelle:1978ye, Sohnius:1981tp}. The Ferrara--Zumino multiplet \cite{Ferrara:1974pz} is dual to the so called old-minimal off-shell formulation of supergravity \cite{Ferrara:1978em, Stelle:1978ye,Komargodski:2010rb}, the $R$-multiplet to the so called new-minimal formulation \cite{Sohnius:1981tp}, and the general current multiplet  to non-minimal formulations of supergravity  \cite{Siegel:1978mj}. 
The field content of the supercurrent superfields determines in this way (the irreducible part of) the auxiliary fields of the different off-shell formulations of supergravity.\footnote{Non-minimal off-shell supergravity defines a reducible multiplet and the general supecurrent superfield only couples to the $16{+}16$ irreducible part.}

The supercurrent superfield can thus be used to probe the leading-order coupling of rigid multiplets to supergravity \cite{Komargodski:2010rb,Festuccia:2011ws}, without relying, in particular, on any specific lagrangian. In this paper we  construct the supercurrent superfield for the massive spin-2 multiplet. The supersymmetry transformations characterising this multiplet actually enjoy a $\mathrm{U}(1)$ $R$-symmetry \cite{Buchbinder:2002gh,Zinoviev:2002xn}, and so it is natural to define the supercurrent superfield as an $R$-multiplet \cite{Sohnius:1981tp}. The resulting supercurrent defines then the coupling of the massive spin-2 multiplet to supergravity in the new-minimal off-shell formulation \cite{Sohnius:1981tp}. We find that the currents entering this $R$-multiplet, and in particular the associated stress-energy tensor, contain improvement terms, namely deformations of the currents that are trivially conserved and hence do not spoil the conserved charges \cite{Callan:1970ze}, some of which are higher-derivative operators. While improvement terms are standard ambiguities of conserved currents in rigid theories -- and their presence in the current multiplets is actually expected \cite{Ferrara:1974pz,Komargodski:2010rb} -- in local theories, where the currents become sources of the associated gauge fields in the lagrangian, the improvement terms acquire physical significance as non-minimal couplings to these gauge fields. Improvements to the stress-energy tensor thus correspond to non-minimal couplings to the metric \cite{Callan:1970ze}. We impose that the improvement terms can be interpreted as couplings to the Riemann tensor by consistency with diffeomorphism invariance. Restricting the analysis to operators with at most 4-derivatives -- which is the order probed by the minimal $R$-multiplet -- we find that only a single coupling satisfies this consistency requirement if we impose an $R$-symmetry multiplet. It is characterised by non-minimal couplings to the Rarita--Schwinger field strength and to the Riemann tensor including higher-derivative operators such as \eqref{RFF}. The resulting coupling to gravity of the massive spin-2 field matches precisely the one arising from the first oscillator mode of the open superstring \cite{Lust:2021jps}. This confirms the expectation that the superstring oscillator modes couple to undeformed supergravity. Combined with the causality issue mentioned above, this provides evidence in favour of the string lamppost principle in four dimensions with only minimal supersymmetry.

We compute the $2\to2$, graviton-mediated elastic scattering amplitude for the massive spin-2 field in the absence of self-couplings between the massive fields. We find that it grows as $s^4$ in both the Regge limit and the high energy limit at fixed scattering angle. Further (on-shell vanishing) couplings of the massive spin-2 multiplet to off-shell supergravity can be written by adding a supersymmetric non-minimal coupling to a real scalar superfield that can be eliminated by a field redefinition. Any such modification gives an amplitude that grows as $s^5$, so that the new-minimal coupling seems to be favoured from this point of view.

The paper is organised as follows. In Section \ref{sec:sc} we introduce the massive spin-2 multiplet and construct the associated supercurrent superfield starting from the $R$-symmetry current. In Section \ref{sec:ms2_sugra} we analyse the coupling to supergravity of the massive spin-2 multiplet defined by this supercurrent, we discuss its inconsistency with respect to diffeomorphism invariance and find that this singles out a unique coupling at the 4-derivative order. In Section \ref{sec:amp} we compute the $2\to2$ elastic scattering amplitude, while in Section \ref{sec:stuckelberg} we discuss how the Kaluza--Klein reduction of pure 5D supergravity requires instead a deformation of the local supersymmetry algebra and gives a different coupling to gravity. We then conclude in Section \ref{sec:conclusions}. Notations and conventions used throughout the paper are collected in Appendix \ref{app:nc}. In Appendix \ref{app:kk}, we describe how to derive the transformation laws of the massive spin-2 supermultiplet from an appropriately chosen Kaluza–Klein reduction of five-dimensional pure supergravity. Finally, Appendix \ref{app:Dj} contains the detailed derivation of the improvement to the axial current used to determine the unique consistent coupling of the massive spin-2 multiplet to supergravity.


\newpage
\section{The massive spin-2 multiplet and the supercurrent superfield} \label{sec:sc}

The first element of our construction is the $D=4$, $\N=1$ rigid supermultiplet of the massive spin-2 field. This multiplet contains one massive spin-2, one massive spin-1 and two Majorana massive spin-$\frac32$ fields, all of mass $m$, which we denote respectively by $h_{\mu\nu}$, $A_\mu$, $\mpsi_\mu$ and $\chi_\mu$. The linearised supersymmetry transformations describing this multiplet are
\begin{equation}
\begin{aligned}
&\delta A_\mu=\frac{1}{2}\bepsilon\mpsi_\mu\; , \\
&\delta\mpsi_\mu=-\frac{i}{4}\(mh_{\mu\rho}+\tilde{F}_{\mu\rho}+2i\partial_\rho A_\mu\gamma_5\)\gamma_5\gamma^\rho\epsilon,  \\
&\delta h_{\mu\nu}=\frac{i}{m}\bepsilon\gamma_5\(\partial_{(\mu}\mpsi_{\nu)}-m\gamma_{(\mu}\chi_{\nu)}\)\; , \\
&\delta\chi_\mu=\frac{i}{4}\partial_\rho h_{\mu\sigma}\gamma^{\rho\sigma}\gamma_5\epsilon+\frac{1}{8m}\partial_\mu F_{\rho\sigma}\gamma^{\rho\sigma}\epsilon+\frac{m}{4}A_\rho\(2\delta^\rho_\mu+\gamma^\rho{}_\mu\)\epsilon\; ,
\end{aligned}
\label{eqn:ms2_susy}
\end{equation}
where $\tilde{F}_{\mu\nu}\equiv\frac12\varepsilon_{\mu\nu\rho\sigma}F^{\rho\sigma}$. These transformations leave invariant the free lagrangian of the massive multiplet, which is the supersymmetric generalisation of the Fierz--Pauli action:\footnote{We choose for convenience a non-canonical normalisation for the bosonic fields. The difference with respect to the canonical one is given by a factor $\sqrt{\frac{2}{3}}$ for the spin-1 field $A_\mu$, and by a factor $\frac14$ for the spin-2 field $h_{\mu\nu}$.}
\beq
\begin{aligned}
	\mathcal{L}_\textup{SFP}=&-\frac{3}{8}F^{\mu\nu}F_{\mu\nu}-\frac{3}{4}m^2A^{\mu}A_\mu -\frac{1}{2}\bar{\mpsi}_\mu\gamma^{\mu\nu\rho}\partial_\nu\mpsi_\rho-\frac{1}{2}\bchi_\mu\gamma^{\mu\nu\rho}\partial_\nu\chi_\rho-m\bar{\mpsi}_\mu\gamma^{\mu\nu}\chi_\nu\\
			&+\frac{1}{4}\partial_\mu h_{\nu\rho}\partial^\nu h^{\mu\rho}-\frac18\partial_\rho h_{\mu\nu}\partial^\rho h^{\mu\nu}-\frac14\partial_\mu h^{\mu\nu}\partial_\nu h+\frac18\partial_\mu h\partial^\mu h-\frac{m^2}{8}\(h^{\mu\nu}h_{\mu\nu}-h^2\) ,
\label{eqn:L_SFP}
\end{aligned}
\eeq

\noindent where we employ the standard definition $h\equiv h^\mu{}_\mu$. The equations of motion that follow from this lagrangian are
\beq
\begin{aligned}
	\begin{aligned}  \partial_\mu F^{\mu\nu}-m^2 A^\nu&=0,  &  \partial_{[\mu}F_{\nu\rho]}&=0, \\
	\gamma^{\mu\nu\rho}\partial_\nu\mpsi_\rho+m\gamma^{\mu\nu}\chi_\nu&=0,  &  \gamma^{\mu\nu\rho}\partial_\nu\chi_\rho+m\gamma^{\mu\nu}\mpsi_\nu&=0, \end{aligned} \\
	 \Box h_{\mu\nu}-2\partial_\rho\partial_{(\mu} h^\rho{}_{\nu)}+\eta_{\mu\nu}\partial_\rho\partial_\sigma h^{\rho\sigma}+\partial_\mu\partial_\nu h-\eta_{\mu\nu}\Box h-m^2\(h_{\mu\nu}-\eta_{\mu\nu }h\)=0, 
\label{eqn:eom_full}
\end{aligned}
\eeq

\noindent and they can be reorganised in terms of the following set of dynamical equations and constraints:
\beq
\begin{aligned}
								 &		& 	&\Box A_\mu =m^2 A_\mu,  		&	 \partial^\mu A_\mu &=0,  		& 	& \\
								 &		& 	&\Box h_{\mu\nu}=m^2 h_{\mu\nu}, 	&	 \partial^\rho h_{\rho\mu}&=0, 	&  	h&=0, \\
	\cancel{\partial}\mpsi_\mu&=m\chi_\mu,  		& 	&\Box\mpsi_\mu=m^2\mpsi_\mu, 	& 	\partial^\mu\mpsi_\mu&=0,	& 	 \gamma^\mu \mpsi_\mu&=0,    \\
	 \cancel{\partial}\chi_\mu&=m\mpsi_\mu, 		& 	&\Box\chi_\mu=m^2\chi_\mu,  		&	 \partial^\mu\chi_\mu&=0,		& 	\gamma^\mu \chi_\mu&=0.    \\
\label{eqn:eom}
\end{aligned}
\eeq

\noindent Upon use of these equations of motion, the transformations \eqref{eqn:ms2_susy} close on
\begin{equation}
	\[\delta_1,\delta_2\]=-\frac{1}{2}\(\bepsilon_1\gamma^\rho\epsilon_2\) \partial_\rho \ .
\label{eqn:susy_cl}
\end{equation}

The massive spin-2 supermultiplet was constructed in the superspace formalism in \cite{Buchbinder:2002gh,Gates:2013tka}, while the transformations \eqref{eqn:ms2_susy} were determined in \cite{Zinoviev:2002xn} by requiring the invariance of the lagrangian \eqref{eqn:L_SFP}. An alternative way to derive this multiplet is by means of a suitable dimensional reduction of pure 5D supergravity. Compactifying this theory on the orbifold $S^1/\mathbb{Z}_2$ breaks supersymmetry from $\N=2$ to $\N=1$ and the 5D field content is projected in four dimensions to pure 4D supergravity coupled to a massless chiral multiplet and a tower of massive spin-2 multiplets. At the linearised level, the supersymmetry transformations of each massive Kaluza--Klein mode decouple from the rest of the fields, resulting in the transformations \eqref{eqn:ms2_susy}. We perform this compactification explicitly in Appendix \ref{app:kk}.\footnote{For related discussions, see \cite{Zinoviev:2002xn, Ondo:2016cdv}.} A key point of this procedure is that the tower of massive spin-2 multiplets comes in the Kaluza--Klein reduction in the so-called St\"uckelberg formulation. The St\"uckelberg gauge symmetry appears in fact naturally in the context of Kaluza--Klein reductions as massive modes of the higher dimensional gauge symmetries.\footnote{For a review on the St\"uckelberg formalism, see for example \cite{Hinterbichler:2011tt} and \cite{Gherghetta:2002nr} for bosonic and fermionic gauge fields respectively.} The parameterisation of Eq.~\eqref{eqn:ms2_susy} corresponds then to the unitary gauge for this St\"uckelberg gauge symmetry. Imposing this unitary gauge is actually a nontrivial step, to which we come back in Section \ref{sec:stuckelberg}.

The aim of this work is to determine the coupling to supergravity of the massive spin-2 supermultiplet defined by Eq.~\eqref{eqn:ms2_susy} in the linearised approximation. At leading order in the Planck mass $M_{\scalebox{0.6}{P}}$, these are the three-point couplings of two massive fields to the metric and the gravitino, which are given respectively by the stress-energy tensor $T^{\mu\nu}$ and the supersymmetry current $\Xi^\mu$. These two currents stand themselves in a supersymmetric multiplet, the so called supercurrent multiplet \cite{Ferrara:1974pz, Sohnius:1981tp, Komargodski:2010rb}. In its most general form the supermultiplet is defined by a vector superfield $\S^{\mu}$ that includes 16+16 components \cite{Komargodski:2010rb}. If the theory admits an $R$-symmetry, one can define the vector superfield $\S^\mu$ such that its first component is the $R$-symmetry current $J^\mu$ and $\S^\mu$ is then a divergence-free linear superfield describing the supercurrent multiplet,  {\it i.e.}
\beq
	D^\alpha D_\alpha \S^\mu = 0 \ , \qquad \partial_\mu \S^\mu = 0 \; , 
\label{eqn:linear_m_ss}
\eeq

\noindent with $D_\alpha$ the superspace covariant derivative. One can write equivalently the linear multiplet constraint using the supersymmetry transformation $\delta_\textup{L}$ involving the left-handed spinor $\epsilon_\textup{L}$, as
\beq
	\delta_\textup{L}^2 J^\mu =0\; ,
\label{eqn:linear_m_comp}
\eeq
using the fact that it factorises into the singlet $\bar \epsilon_\textup{L} \epsilon_\textup{L}$.\footnote{For an explicit application of the linear superfield constraint in components, see Section \ref{sec:ms2_sugra} and the related Appendix \ref{app:Dj}.} Together with $J^\mu$, $\Xi^\mu$ and $T^{\mu\nu}$, the multiplet includes a trivially conserved antisymmetric tensor $X^{\mu\nu}$, which can be written as the dual of a curl $X^{\mu\nu}=\varepsilon^{\mu\nu\rho\sigma}\partial_\rho X_\sigma$. The rigid supersymmetry transformations of this supercurrent multiplet are given by \cite{Sohnius:1981tp}
\beq
\begin{aligned}
&\delta J_\mu=-i\bar{\epsilon}\gamma_5\Xi_\mu, \\
&\delta \Xi_\mu=  T_{\mu\nu}\gamma^\nu\epsilon+X_{\mu\nu}\gamma^\nu\epsilon+\frac{1}{8}\varepsilon_{\mu\nu\rho\sigma}\partial^\rho J^\sigma\gamma^\nu \epsilon+\frac{i}{4}\partial_\rho J_\mu\gamma_5\gamma^\rho\epsilon, \\
&\delta T_{\mu\nu}=-\frac{1}{4}\bepsilon\gamma_{(\mu}{}^{\lambda}\partial_\lambda \Xi_{\nu)}, \\
&\delta X_{\mu}=\frac{i}{8}\bepsilon\gamma_5\gamma^\mu\gamma^\lambda \Xi_\lambda.  
\end{aligned}
\label{eqn:sc_susy}
\eeq

An important feature of the supercurrent superfield is that, being a multiplet of currents, it is in general subject to improvements \cite{Ferrara:1974pz, Komargodski:2010rb, Callan:1970ze}, namely deformations which are trivially conserved and thus do not modify the conserved charges. While in rigid theories the improvements are standard deformations of the conserved currents, in the local case they correspond to non-minimal couplings to the massless fields \cite{Callan:1970ze} and acquire therefore a physical meaning. Such non-minimal couplings will play a crucial role in the analysis that follows.

The first step is to determine an example of supercurrent superfield for the massive spin-2 supermultiplet, which we now construct explicitly following Eq.~\eqref{eqn:sc_susy}. The transformations \eqref{eqn:ms2_susy} possess the following $R$-symmetry:
\begin{align}
    \mpsi_\mu \to e^{i \alpha \gamma_5}\mpsi_\mu\; , && \chi_\mu\to e^{-i \alpha \gamma_5}\chi_\mu\; , && \epsilon\to e^{-i \alpha \gamma_5}\epsilon\; ,
\label{eqn:axial_sym}
\end{align}

\noindent with $\alpha$ being the $R$-symmetry $\mathrm{U}(1)$ parameter. The simplest axial current associated to this symmetry is obtained from the Noether procedure as
\begin{equation}
J^\mu_0=\varepsilon^{\mu\nu\rho\sigma}\(\bar{\mpsi}_\rho\gamma_\nu\mpsi_\sigma-\bchi_\rho\gamma_\nu\chi_\sigma\)\; .
\label{eqn:axial_c}
\end{equation}

\noindent One can check that this current is conserved on-shell and that it does define a linear multiplet according to Eq.~\eqref{eqn:linear_m_comp}. Thus, the axial current $J^\mu_0$ is the natural starting point from which to construct the supercurrent multiplet for the massive spin-2 superfield, which we denote as $\S^\mu_0$. This current is unique as a dimension-three operator, but we will see in Section \ref{sec:ms2_sugra} that one needs to introduce dimension-four improvement terms in order to define a consistent coupling to supergravity. Acting on $J^\mu_0$ with a supersymmetry transformation \eqref{eqn:ms2_susy} yields, according to (\ref{eqn:sc_susy}), the following current of supersymmetry
\begin{equation}
\begin{aligned}
\Xi^\mu_0=\frac{1}{2}\varepsilon^{\mu\nu\rho\sigma}&\biggl\{\(mh_{\rho\tau}+\tilde{F}_{\rho\tau}-2i\partial_\tau A_\rho\gamma_5\)\gamma^\tau\gamma_\nu\mpsi_\sigma \\
&\quad+\[\partial_\lambda h_{\tau\rho}\gamma^{\lambda\tau}\gamma_5-\frac{i}{2m}\partial_\rho F_{\lambda\tau}\gamma^{\lambda\tau}+imA_\tau\(2\delta^\tau_\rho+\gamma_\rho{}^\tau\)\]\gamma_\nu\gamma_5\chi_\sigma\biggr\}\; . 
\end{aligned}
\label{eqn:susy_c}
\end{equation}

\noindent This supersymmetry current is also conserved on-shell by construction. Acting again on $\Xi^\mu_0$ with \eqref{eqn:ms2_susy} provides both the stress-energy tensor $T^{\mu\nu}_0$ and the two-form $X^{\mu\nu}_0$, according to Eq.~\eqref{eqn:sc_susy}. The latter is equal to
\begin{equation}
X^{\mu}_0=\frac{1}{2}A_\nu\(mh^{\nu\mu}+\tilde{F}^{\nu\mu}\)+\frac{i}{4}\bar{\mpsi}^\nu\gamma_5 \gamma^\mu\mpsi_\nu\; ,
\label{eqn:sc_X}
\end{equation}

\noindent while the former takes instead the peculiar form
\beq
T^{\mu\nu}_0=T^{\mu\nu}_\textup{K}+\partial_\rho B^{\rho\mu\nu}_0+\partial_\rho\partial_\sigma G^{\mu\sigma\nu\rho}_0,
\label{eqn:sc_T_1}
\eeq

\noindent which we now analyse in detail. The term $T^{\mu\nu}_\textup{K}$ is the standard contribution coming from the kinetic terms of the various fields, namely those characterising the Proca, Rarita--Schwinger and Fierz--Pauli lagrangians:
\bea
T^{\mu\nu}_\textup{K}&=&\frac{3}{2}T^{\mu\nu}_\textup{P}+T^{\mu\nu}_\textup{RS}+\frac{1}{4}T^{\mu\nu}_\textup{FP}\\
	&=&\frac{3}{2}\(F^{\mu\rho}F^\nu{}_\rho-\frac{1}{4}\eta^{\mu\nu}F^2+m^2A^\mu A^\nu-\frac{m^2}{2}\eta^{\mu\nu}A^2\) \nonumber \\
	&&+\frac{1}{2}\(\bar{\mpsi}^\sigma\gamma^{(\mu}\partial^{\nu)}\mpsi_\sigma+\bchi^\sigma\gamma^{(\mu}\partial^{\nu)}\chi_\sigma\)-\(\bar{\mpsi}^\sigma\gamma^{(\mu}\partial_\sigma\psi^{\nu)}+\bchi^\sigma\gamma^{(\mu}\partial_\sigma\chi^{\nu)}\)  \nonumber\\
	&&+\frac{1}{4}\partial^\mu h^{\rho\sigma}\partial^\nu h_{\rho\sigma}+\frac{1}{2}\partial^\rho h^{\mu\sigma}\partial_\rho h^{\nu}{}_{\sigma}-\partial^{(\mu|} h^{\rho\sigma}\partial_\rho h^{|\nu)}{}_{\sigma}+\frac{1}{2}\partial^\rho h^{\mu\sigma}\partial_\sigma h^\nu{}_\rho -\frac{1}{2}h^{\rho\sigma}\partial_\rho\partial_\sigma h^{\mu\nu}\nonumber \\
    && +\frac{1}{2}m^2h^{\mu\rho}h^{\nu}{}_{\rho}+\frac{1}{4}\eta^{\mu\nu}\(-\frac{1}{2}\partial^\lambda h^{\rho\sigma}\partial_\lambda h_{\rho\sigma}+\partial^\lambda h^{\rho\sigma}\partial_\rho h_{\lambda\sigma}-\frac{m^2}{2}h^{\rho\sigma}h_{\rho\sigma}\)\; . \nonumber
\label{eqn:sc_T_K}
\eea

\noindent These are the standard quadratic symmetric Hilbert stress-energy tensors evaluated on the Minkowski background.\footnote{We refer to \cite{Casagrande:2023fjk} and \cite{Petrov:2017bdx} for the full covariant stress-energy tensors for the Rarita--Schwinger and Fierz--Pauli actions respectively.} Note that the resulting tensor for the Rarita--Schwinger field differs from the one obtained directly in flat space (namely from \eqref{eqn:L_SFP} via the Belinfante procedure) by terms which come from the connections of the original covariant derivatives. Despite being improvements from the flat space point of view, these terms are crucial in order to reconstruct a consistent coupling to gravity of the kinetic terms and thus for a correct interpretation of the stress-energy tensor entering the supercurrent superfield $\S^\mu_0$ under consideration.

The terms $\partial_\rho B^{\rho\mu\nu}_0$ and $\partial_\rho\partial_\sigma G^{\mu\sigma\nu\rho}_0$ are also improvements with respect to $T^{\mu\nu}_\textup{K}$ \cite{Callan:1970ze}. They possess the following symmetries:
\begin{equation}
\begin{aligned}
\partial_\rho B^{\rho\mu\nu}_0=\partial_\rho B^{\rho(\mu\nu)}_0=\partial_\rho B^{[\rho\mu]\nu}_0=\partial_\rho B^{[\rho|\mu|\nu]}_0,&\\
\partial_\rho\partial_\sigma G^{\mu\sigma\nu\rho}_0= \partial_\rho\partial_\sigma G^{(\mu|\sigma|\nu)\rho}_0=\partial_\rho\partial_\sigma G^{[\mu\sigma]\nu\rho}_0=\partial_\rho\partial_\sigma G^{\mu\sigma[\nu\rho]}_0, &
\end{aligned}
\label{eqn:imp_sym}
\end{equation}

\noindent which imply that they are trivially conserved. Their explicit expressions are, respectively,
\begin{equation}
\begin{aligned}
B^{\rho\mu\nu}_0=&\frac{1}{4}\(h^{(\nu}{}_{\sigma}\partial^{\mu)}h^{\rho\sigma}-\partial^{(\mu}h^{\nu)}{}_{\sigma}h^{\rho\sigma}\)+\frac{5}{4}\(A^{(\nu}\partial^{\mu)} A^\rho-A^\rho\partial^{(\mu}A^{\nu)}\) \\
&  +\frac{1}{4m}\(\partial^{(\mu}\tilde{F}^{\rho\sigma}h^{\nu)}{}_\sigma-\tilde{F}^{\rho\sigma}\partial^{(\mu}h^{\nu)}{}_\sigma+\tilde{F}^{(\nu}{}_\sigma\partial^{\mu)}h^{\rho\sigma}-\partial^{(\mu}\tilde{F}^{\nu)}{}_\sigma h^{\rho\sigma}\)\\
& +\frac{1}{2m}\(\partial^{(\mu}\bchi^{\nu)}\mpsi^\rho-\bchi^{(\nu}\partial^{\mu)}\mpsi^\rho+\partial^{(\mu}\bar{\mpsi}^{\nu)}\chi^\rho-\bar{\mpsi}^{(\nu}\partial^{\mu)}\chi^\rho\)\; , 
\end{aligned}
\label{eqn:sc_B}
\end{equation}

\noindent and \footnote{Throughout the paper we employ the following notation on symmetrised indices: $2\eta^{\nu][\sigma}F^{\mu]\lambda}F_{\lambda}{}^{[\rho} \equiv \eta^{\nu[\sigma}F^{\mu]\lambda}F_{\lambda}{}^{\rho}-\eta^{\rho[\sigma}F^{\mu]\lambda}F_{\lambda}{}^{\nu} $. See Appendix \ref{app:nc} for more details.}
\beq
\begin{aligned}
    G^{\mu\sigma\nu\rho}_0=&\frac{1}{2m^2}\eta^{\nu][\sigma}F^{\mu]\lambda}F_{\lambda}{}^{[\rho}+\frac{1}{8m^2}\eta^{\nu[\sigma}\eta^{\mu]\rho}F^2+3\eta^{\nu][\sigma}A^{\mu]}A^{[\rho}+\frac{1}{4}\eta^{\nu[\mu}\eta^{\sigma]\rho}A^2\\
    &+\frac{1}{2m}\eta^{\rho][\mu}\(\tilde{F}^{\sigma]}{}_{\lambda}h^{\lambda[\nu}-h^{\sigma]}{}_{\lambda}\tilde{F}^{\lambda[\nu}\)-h^{\mu[\rho}h^{\nu]\sigma}+\frac{1}{2}\eta^{\rho][\mu}h^{\sigma]}{}_\lambda h^{\lambda[\nu}\\
    &+\frac{1}{m}\(\bar{\mpsi}^{[\sigma}\gamma^{\mu][\rho}\chi^{\nu]}+\bar{\mpsi}^{[\rho}\gamma^{\nu][\sigma}\chi^{\mu]}\)\; . 
\end{aligned}
\label{eqn:sc_G}
\eeq
While the symmetry properties (\ref{eqn:imp_sym}) are manifest for $G^{\mu\sigma\nu\rho}_0$ in \eqref{eqn:sc_G}, those of $B^{\rho\mu\nu}_0$ require to use the equations of motion.\footnote{Taking the Proca sector of Eq.~(\ref{eqn:sc_B}) as an example, we have in fact 
\begin{equation}
\begin{aligned}
\frac{4}{5}\partial_\rho B_A^{\rho\mu\nu}=&\partial_\rho \(A^{(\nu}\partial^{\mu)} A^\rho-A^\rho\partial^{(\mu}A^{\nu)}\)=\\
=&\partial_\rho \(A^\nu\partial^{[\mu}A^{\rho]}+A^{[\mu}\partial^{\rho]}A^\nu+A^{[\mu|}\partial^\nu A^{|\rho]}\)+\underbrace{\partial_\rho\(A^{[\nu|}\partial^\rho A^{|\mu]}\)}_{=0}=\frac{4}{5}\partial_\rho B_A^{[\rho\mu]\nu}\; . 
\end{aligned}
\end{equation}

\noindent and analogous manipulations yield also $\partial_\rho B^{\rho\mu\nu}=\partial_\rho B^{[\rho|\mu|\nu]}$.} The presence of improvements in the currents of the multiplet is rather standard \cite{Ferrara:1974pz} and they are perfectly consistent from the rigid theory point of view. The peculiar property of the improvements \eqref{eqn:sc_B}-\eqref{eqn:sc_G} is that they contain higher-derivative operators, which are required by supersymmetry.

Thus, we determined the supercurrent superfield associated to the massive spin-2 supermultiplet \eqref{eqn:ms2_susy}, which admits the components 
\beq
	\S^\mu_0=\left\{J^\mu_0,\Xi^\mu_0,T^{\mu\nu}_0,X^{\mu\nu}_0\right\},
\label{eqn:sc_0}
\eeq

\noindent with $J^\mu_0$ being the $R$-symmetry current \eqref{eqn:axial_c} and the other currents defined respectively in \eqref{eqn:susy_c}, \eqref{eqn:sc_T_1} and \eqref{eqn:sc_X}. These currents are related by the supersymmetry transformations \eqref{eqn:sc_susy}, which close on-shell on \eqref{eqn:susy_cl}.

\section{The coupling to supergravity and its uniqueness} \label{sec:ms2_sugra}

The supercurrent superfield $\S^\mu_0$ we determined in the previous section can be used to define the coupling of the rigid multiplet $\left\{A_\mu,\mpsi_\mu,h_{\mu\nu},\chi_\mu\right\}$ to supergravity at leading order in $M_{\scalebox{0.6}{P}}$. In going from global to local supersymmetry, the supercurrent superfield has in fact the natural interpretation of being the multiplet of currents associated to the gauge symmetries of supergravity. 
The supercurrent superfields actually define a coupling to off-shell supergravity. The current of supersymmetry and the stress-energy tensor couple respectively to the gravitino and to the fluctuation of the metric, while the other currents of the multiplet couple to the auxiliary fields of the supergravity multiplet. Different types of supercurrent supermultiplets are then associated with different off-shell formulations of supergravity \cite{Ferrara:1978em, Stelle:1978ye, Sohnius:1981tp,Siegel:1978mj}. In particular, the $R$-symmetry multiplet \eqref{eqn:sc_susy} is associated to the so called new-minimal off-shell supergravity \cite{Sohnius:1981tp}, in which the auxiliary fields, that couple respectively to the $R$-symmetry current $J^\mu$ and the antisymmetric field $X^{\mu\nu}$, are an axial gauge field $V^\mu$ and an antisymmetric tensor field $C_{\mu\nu}$ of field strength $G^\mu =\frac12  \varepsilon^{\mu\nu\rho\sigma} \partial_\nu C_{\rho\sigma} $. The linearised coupling lagrangian of new-minimal off-shell supergravity takes the form\footnote{Here we have defined the axial gauge field as it appears in conformal supergravity, whereas \cite{Sohnius:1981tp} uses the vector $A^{\scalebox{0.5}{SW}}_\mu = V_\mu - \frac38 C_\mu$.} \cite{Sohnius:1981tp}
\begin{equation}
\mathcal{L}\sim-\frac{1}{2}(g_{\mu\nu}-\eta_{\mu\nu})T^{\mu\nu}-\frac12 \bar{\psi}_\mu \Xi^\mu - \frac12 V_\mu J^\mu + G_\mu X^\mu \;,
\label{eqn:new_min_min_coupling}
\end{equation}

\noindent and the local supersymmetry transformations of the off-shell supergravity multiplet are
\beq
\begin{aligned}
&\delta g_{\mu\nu} = \bar \epsilon \gamma_{(\mu} \psi_{\nu)}\; , \\
&\delta \psi_\mu =  \partial_\mu \epsilon - \frac14 \partial_\nu g_{\rho\mu} \gamma^{\nu\rho} \epsilon + i \gamma_5 V_\mu \epsilon - \frac{i}{4} \gamma_\mu G_\nu \gamma_5 \gamma^\nu \epsilon \; ,  \\
&\delta V_\mu = - \frac{i}{4} \bar \epsilon \gamma_5 \gamma_\mu \gamma^{\nu \rho} \partial_\nu \psi_\rho + \frac{3}{8} \varepsilon_{\mu\nu}{}^{\rho\sigma} \bar \epsilon \gamma^\nu \partial_\rho \psi_\sigma \; , \\
&\delta C_{\mu\nu} =\bar\epsilon \gamma_{[\mu} \psi_{\nu]} \; .
\end{aligned}
\label{eqn:nm_susy}
\eeq

\noindent One can check that the set of supersymmetry transformations given by \eqref{eqn:sc_susy} and \eqref{eqn:nm_susy} leave the lagrangian \eqref{eqn:new_min_min_coupling} invariant.

Here we should stress that the energy-momentum tensor \eqref{eqn:sc_T_1} is on-shell for the linearised equations of motion of the massive fields \eqref{eqn:eom}. This is very natural for a stress-energy tensor as the conservation only holds on-shell. However, it follows that the stress-tensor only determines the three-point couplings \eqref{eqn:new_min_min_coupling}  in the lagrangian modulo terms that vanish using the linearised equations of motion of the massive free field.  Nonetheless, such terms can always be reabsorbed at leading order by field redefinitions linear in the massive fields such that they only introduce higher order terms in the massless fields. Some of these terms follow straightforwardly by covariantisation of the lagrangian under (super-)diffeomorphisms, and they should be computed to obtain the non-linear lagrangian invariant under local supersymmetry, but they do not play any role in the analysis below and in the 2-to-2 scattering amplitude we compute in the next section.

While the improvements represent ambiguities in  the definition of the conserved currents that do not affect the conserved charges when they are associated to rigid symmetries, 
 they acquire physical meaning and are understood as non-minimal couplings to the associated massless gauge fields when these symmetries are gauged  \cite{Callan:1970ze}. Focusing on the stress-energy tensor $T^{\mu\nu}_0$ entering the supercurrent superfield $\S^\mu_0$ \eqref{eqn:sc_0}, and in particular on its decomposition \eqref{eqn:sc_T_1}, the term $T^{\mu\nu}_\textup{K}$ clearly results, as stated before, from the minimal coupling to gravity obtained by covariantising the free lagrangian \eqref{eqn:L_SFP} and taking the symmetric Hilbert stress-energy tensor.\footnote{Note that if one redefines the lagrangian up to a total derivative in flat spacetime, one can also introduce a different lagrangian minimally coupled to gravity in curved spacetime resulting in an ambiguity in the definition of $T^{\mu\nu}_\textup{K}$. The canonical choice is fixed by the covariantisation of the massless theory, in particular the definition of the Pauli--Fierz lagrangian in curved spacetime is determined from linearised gravity around a solution to Einstein equations  \cite{Petrov:2017bdx}.} The improvement terms $B^{\rho\mu\nu}_0$ in \eqref{eqn:sc_B} and $G^{\mu\sigma\nu\rho}_0$ in \eqref{eqn:sc_G} correspond then to non-minimal couplings to the metric field. Their specific form is determined by the different symmetries  \eqref{eqn:imp_sym} of the two type of improvements. Looking at the linearised coupling of the stress-energy tensor to the metric fluctuation $g_{\mu\nu}$ in Eq.~\eqref{eqn:new_min_min_coupling}, one can see that $G^{\mu\sigma\nu\rho}_0$ defines a non-minimal coupling to the linearised Riemann tensor:
\begin{equation}
    g_{\mu\nu}\partial_\rho\partial_\sigma G^{\mu\sigma\nu\rho}_0\approx\(\partial_{\rho]}\partial_{[\sigma} g_{\mu][\nu}\)G^{[\mu\sigma][\nu\rho]}_0\sim   - \frac12  R_{\mu\sigma\nu\rho}G^{\mu\sigma\nu\rho}_0\, ,
\label{eqn:riem_L}
\end{equation}

\noindent while $B^{\rho\mu\nu}_0$ corresponds to a non-minimal coupling to the linearised Levi--Civita connection:
\begin{equation}
    g_{\mu\nu}\partial_\rho B^{\rho\mu\nu}_0\approx\(\partial_{[\mu} g_{\rho]\nu}+\frac{1}{2}\partial_\nu g_{\mu\rho}\)B^{[\rho\mu]\nu}_0\sim\Gamma_{\rho\mu\nu}B^{\rho\mu\nu}_0,
\label{eqn:lc_L}
\end{equation}

\noindent where the linearised Riemann tensor and Levi--Civita connection are, respectively,
\begin{align}
R_{\mu\sigma\nu\rho}\sim  -2 \partial_{\rho]}\partial_{[\sigma}g_{\mu][\nu}\; , && g_{\rho\sigma} \Gamma^\sigma_{\mu\nu}\sim\partial_{[\mu}g_{\rho]\nu}+\frac{1}{2}\partial_\nu g_{\mu\rho}\; .
\label{eqn:riem_lc_lin}
\end{align}

\noindent The couplings \eqref{eqn:riem_L} to the Riemann tensor associated to the improvements $G^{\mu\sigma\nu\rho}_0$ \eqref{eqn:sc_G}, despite being non-standard because of the higher-derivative terms, are perfectly fine from the point of view of diffeomorphism invariance. On the other hand, the improvements $B^{\rho\mu\nu}_0$ \eqref{eqn:sc_B} represent couplings to the bare Levi--Civita connection \eqref{eqn:lc_L}, \textit{i.e.} not embedded in any covariant derivatives, and are therefore not consistent with diffeomorphism invariance. Hence, because of the presence of such non-diffeomorphism invariant $B^{\rho\mu\nu}_0$-terms, the supercurrent superfield $\S^\mu_0$ of Eq.~\eqref{eqn:sc_0} does not define a consistent coupling to supergravity of the massive spin-2 multiplet \eqref{eqn:ms2_susy}.




Thus, although the supercurrent $\S^\mu_0$ of Eq.~\eqref{eqn:sc_0} is inherently associated with the massive spin-2 multiplet \eqref{eqn:ms2_susy}, as it arises from its $R$-symmetry Noether current \eqref{eqn:axial_c}, we find that we must correct it by an improvement superfield in order to define its coupling to supergravity. The most general improvement superfield can be written in superspace as
\beq  \S^\mu_0 \ \longrightarrow \ \S^\mu=\S^\mu_0+\partial_\nu \mathcal{L}^{\mu\nu} + \sigma^{\mu\, \alpha\dot{\beta}} [ D_\alpha , \bar D_{\dot{\beta}} ] \mathcal{U}  \label{GeneralImproved} \eeq
where $\mathcal{U}$ is a real unconstrained superfield and $\mathcal{L}^{\mu\nu}  =- \mathcal{L}^{\nu\mu}$ is a linear superfield, \textit{i.e.} $D^\alpha D_\alpha \mathcal{L}^{\mu\nu}=0$. For convenience we define the additional components accordingly as
\begin{align}
  \Delta\S^\mu=\left\{\Delta J^\mu,\Delta\Xi^\mu,\Delta T^{\mu\nu},\Delta X^{\mu\nu}\right\}.
\label{eqn:DJ_def}
\end{align}
Introducing a non-trivial superfield improvement $\mathcal{U}$ breaks the conservation of the first component of $\S^\mu$ and gives a general supercurrent multiplet. This more general case would appear in coupling the massive spin-2 multiplet to non-minimal supergravity \cite{Siegel:1978mj}. In this section we consider the most general supercurrent superfield and will determine the ones -- if any -- that define a consistent coupling to off-shell supergravity, imposing that the non-minimal couplings be diffeomorphism invariant. This condition corresponds to requiring the absence of $B^{\rho\mu\nu}_0$-like improvement terms in the resulting stress-energy tensor.

Before performing this analysis, we should emphasise that this procedure does not define the most general coupling of the spin-2 supermultiplet to supergravity. Assuming that the supercurrent multiplet $\S^\mu$ satisfies the transformation rules \eqref{eqn:sc_susy} is equivalent to imposing that the supersymmetry transformations of the massless fields of the new-minimal multiplet are not modified. We will discuss at the end of this section the generalisation of $\S^\mu$ being a general supercurrent multiplet, corresponding to the coupling of the massive spin-2 multiplet to non-minimal off-shell supergravity, but in general one may expect deformations of the supersymmetry transformations of the massless fields that are not compatible with any off-shell formulation of supergravity.  As we will see in Section \ref{sec:stuckelberg}, this is the case for the Kaluza--Klein theory obtained by compactification on $S^1 / \mathbb{Z}_2$, in which the supersymmetry algebra is modified by higher-derivative terms in the unitary gauge. In this section we shall find instead that there is a unique way to couple the spin-2 multiplet to new-minimal supergravity with terms involving up to four derivatives, and which differs from Kaluza--Klein theory.

We therefore assume here that the local supersymmetry algebra is undeformed and look at the improvement \eqref{GeneralImproved}. We first take $\mathcal{U}=0$ and proceed analogously to the previous section by starting from the simplest element of the improvement multiplet \eqref{eqn:DJ_def}, \textit{i.e.} the axial current $\Delta J^\mu$. This improvement has the general form
\begin{equation}
\Delta J^\mu=\frac{1}{m}\partial_\nu L^{\mu\nu},
\label{eqn:Dj_gen}
\end{equation}

\noindent where $L^{\mu\nu}$ is a real, antisymmetric tensor, which ensures that the improvement is trivially conserved. The most general ansatz for such an improvement that is compatible with the setup under consideration consists of a $\Delta J^\mu$ which is quadratic in the fields, of mass dimension three and has the spacetime parity of an axial vector. Most importantly, the deformation has to be consistent with the linear supermultiplet structure \eqref{eqn:linear_m_comp}, namely
\beq
\delta_\textup{L}^2L^{\mu\nu}=0\; .
\label{eqn:Dj_lin_m}
\eeq

\noindent We also restrict to improvements $\Delta J^\mu$ which contain operators that produce contributions to the stress-energy tensor with at most four derivatives, like the ones found previously in Eq.~\eqref{eqn:sc_T_1}, \eqref{eqn:sc_B} and \eqref{eqn:sc_G}.

We find that the general ansatz for the axial current improvement term satisfying the linear superfield constraint \eqref{eqn:Dj_lin_m} and with at most two derivatives terms can be written as follows:\footnote{More details on how to impose the various requirements and obtain the ansatz \eqref{eqn:Dj_fin} are collected in Appendix \ref{app:Dj}.}
\begin{equation}
	L^{\mu\nu}=c_\textup{B}\(2h^{[\mu|}{}_\lambda\partial^\lambda A^{|\nu]}+\frac{1}{m}\varepsilon^{\mu\nu\rho\sigma}\partial_\lambda A_\rho\partial_\sigma A^\lambda\)+c_\textup{F1}\varepsilon^{\mu\nu\rho\sigma}\bchi_\rho\mpsi_\sigma+ic_\textup{F2}\bchi^{[\mu}\gamma_5\mpsi^{\nu]}.
\label{eqn:Dj_fin}
\end{equation}

\noindent In order to find a consistent coupling to new-minimal supergravity, we have to fix the three coefficients $\left\{c_\textup{B},c_\textup{F1},c_\textup{F2}\right\}$ such that the improvement term of type $ B^{\rho\mu\nu}$ in $T^{\mu\nu}$ vanishes.  The resulting improvement to the stress-energy tensor takes again the form of Eq.~\eqref{eqn:sc_T_1}, namely
\begin{equation}
    \Delta T^{\mu\nu}=\partial_\rho\Delta B^{\rho\mu\nu}+\partial_\rho\partial_\sigma\Delta G^{\mu\sigma\nu\rho},
\end{equation}
The diffeomorphism-invariant couplings are then defined for those coefficients $\left\{c_\textup{B},c_\textup{F1},c_\textup{F2}\right\}$ such that $\Delta B^{\rho\mu\nu}=-B^{\rho\mu\nu}_0$, so that the $B^{\rho\mu\nu}$-like terms cancel exactly in $T^{\mu\nu}_0+\Delta T^{\mu\nu}$. The contributions of each of the three axial improvements of Eq.~\eqref{eqn:Dj_fin} turn out to be
\begin{align}
    &\begin{aligned}
        \partial_\rho\Delta B^{\rho\mu\nu}_\textup{B}=&\frac{1}{8}\partial_\rho\(A^{(\mu}\partial^{\nu)}A^\rho-A^\rho \partial^{(\mu}A^{\nu)}\)\\
        	&+\frac{1}{8m^2}\partial_\rho\(\partial^\sigma A^{(\mu}\partial_\sigma\partial^{\nu)}A^\rho-\partial_\sigma A^\rho \partial^\sigma\partial^{(\mu}A^{\nu)}\),
    \end{aligned} \label{eqn:DB_B} \\
     &\begin{aligned}
        \partial_\rho \Delta B^{\rho\mu\nu}_\textup{F1}=&\frac{1}{2m}\partial_\rho\(\partial^{(\mu}\bchi^{\nu)}\mpsi^\rho-\bchi^{(\nu}\partial^{\mu)}\mpsi^\rho+\partial^{(\mu}\bar{\mpsi}^{\nu)}\chi^\rho-\bar{\mpsi}^{(\nu}\partial^{\mu)}\chi^\rho\)\\
        &+\frac{1}{2m}\partial_\rho\(\partial^{(\mu|}\bar{\mpsi}_\sigma\gamma^\rho\gamma^{|\nu)}\chi^\sigma-\partial^\rho\bar{\mpsi}_\sigma\gamma^{(\mu}\gamma^{\nu)}\chi^\sigma\), 
    \end{aligned} \label{eqn:DB_F1} \\
    &\partial_\rho\Delta B^{\rho\mu\nu}_\textup{F2}=-\frac{1}{4}\partial_\rho B^{\rho\mu\nu}_0. \label{eqn:DB_F2}
\end{align}

\noindent Therefore, the axial improvements dubbed with B and F1 in \eqref{eqn:Dj_fin} lead to \textit{new} non-diffeomorphism invariant improvements to the stress-energy tensor, which appear in the second line of Eq.~\eqref{eqn:DB_B} and \eqref{eqn:DB_F1} respectively, while the improvement F2 is proportional to the non diffeomorphism-invariant improvement previously found in \eqref{eqn:sc_B}. Hence there is a unique choice of the coefficients in \eqref{eqn:Dj_fin}, namely
\be
	c_\textup{B}=c_\textup{F1}=0\;, \qquad c_\textup{F2}=4\; ,
\label{eqn:scu_coeff_sol}
\ee

\noindent that removes the $B^{\rho\mu\nu}$-like terms \eqref{eqn:sc_B} affecting the minimal supercurrent supermultiplet $\S_0^\mu$ without introducing additional terms of the same type.

We conclude from this analysis that there exists a unique three-point coupling, including at most four derivatives, of the massive spin-2 supermultiplet \eqref{eqn:ms2_susy} to new-minimal off-shell supergravity which is consistent both with supersymmetry and diffeomorphism invariance. This unique coupling is defined by the following supercurrent supermultiplet:
\begin{equation}
	\S^\mu=\left\{ J^\mu,\Xi^\mu,T^{\mu\nu},X^{\mu\nu} \right\},
\label{eqn:scu_tot}
\end{equation}

\noindent with 
\bea
J^\mu&=&\varepsilon^{\mu\nu\rho\sigma}\(\bar{\mpsi}_\rho\gamma_\nu\mpsi_\sigma-\bchi_\rho\gamma_\nu\chi_\sigma\)+\frac{4i}{m}\partial_\nu\(\bchi^{[\mu}\gamma_5\mpsi^{\nu]}\)\; ,  \label{eqn:scu_j} \\
    	    \Xi^\mu&=&\Xi^\mu_0+\frac{i}{m}\partial_\nu\biggl[\(mh^{[\nu}{}_\lambda+\tilde{F}^{[\nu}{}_\lambda-2i\partial_\lambda A^{[\nu}\gamma_5\)\gamma^{\lambda}\gamma_5\chi^{\mu]}  \nonumber \\
&&     +\(\partial_\rho h^{[\mu}{}_\sigma\gamma^{\rho\sigma}\gamma_5-\frac{i}{2m}\partial^{[\mu}F_{\rho\sigma}\gamma^{\rho\sigma}+2imA^{[\mu}-imA_\rho\gamma^{\rho[\mu}\)\mpsi^{\nu]}\biggr]\; ,  \label{eqn:scu_Xi} \\
    T^{\mu\nu}&=&T^{\mu\nu}_\textup{K}+\partial_\rho\partial_\sigma \G^{\mu\sigma\nu\rho}\; ,  \label{eqn:scu_T} \\
X^{\mu}&=&\frac{i}{4}\bar{\mpsi}^\nu\gamma_5 \gamma^\mu\mpsi_\nu+\frac{3m}{4}A_{\nu}h^{\nu\mu}+\frac{9}{8}A_{\nu}\tilde{F}^{\nu\mu}+\frac{1}{4m^2}\partial_\rho\partial_\sigma\(\tilde{F}^{\mu\rho}A^\sigma\)  \nonumber \\
    && +\frac{3}{8m}\partial_\nu\(\partial^\mu A^\rho h^\nu{}_\rho+\partial_\rho A^\nu h^{\mu\rho}\)-\frac{1}{8m}\partial_\nu\(\partial^\rho A^\mu h^{\nu}{}_{\rho}+\partial^\nu A_\rho h^{\mu\rho}\)  \\
    && -\frac{1}{8}\varepsilon^{\mu\nu\rho\sigma}\partial_\nu h^\lambda{}_{\rho} h_{\sigma\lambda}+\frac{1}{4m}\varepsilon^{\mu\nu\rho\sigma}\(\bchi_\nu \partial_\rho\mpsi_\sigma+\bar{\mpsi}_\nu \partial_\rho\chi_\sigma\)\; , \nonumber
\label{eqn:scu_X} 
\eea

\noindent where $\Xi_0^\mu$ represents the terms originating from the minimal setup, defined in Eq.~\eqref{eqn:susy_c}, while $T^{\mu\nu}_\textup{K}$ is the Hilbert stress-energy tensor \eqref{eqn:sc_T_K}. The resulting coupling to supergravity is characterised by the higher-derivative, non-minimal coupling to the Riemann tensor defined by 
\begin{equation}
    \begin{aligned}
        \mathcal{G}^{\mu\sigma\nu\rho}=&-\frac{1}{2}h^{\rho[\mu}h^{\sigma]\nu}+\frac{1}{2}\eta^{\rho][\mu}h^{\sigma]}{}_\lambda h^{\lambda[\nu}+2\eta^{\rho][\mu}A^{\sigma]}A^{[\nu}-\frac{3}{4m^2}F^{\mu\sigma}F^{\nu\rho}+\frac{1}{2m^2}\eta^{\rho][\mu}F^{\sigma]}{}_\lambda F^{\lambda[\nu}\\
        &+\frac{1}{m^2}\(\eta^{\sigma][\nu}F^{\rho]}{}_\lambda \partial^\lambda A^{[\mu}+\eta^{\rho][\mu}F^{\sigma]}{}_\lambda \partial^\lambda A^{[\nu}\)+\frac{1}{2m}\eta^{\rho][\mu}\(\tilde{F}^{\sigma]}{}_{\lambda}h^{\lambda[\nu}-h^{\sigma]}{}_{\lambda}\tilde{F}^{\lambda[\nu}\)\\
        &+\frac{1}{2m}\(\varepsilon^{\mu\sigma\lambda\tau} \partial_\lambda h^{[\rho}{}_\tau A^{\nu]}+\varepsilon^{\nu\rho\lambda\tau} \partial_\lambda h^{[\sigma}{}_\tau A^{\mu]}\)+\frac{1}{m}\(\bar{\mpsi}^{[\sigma}\gamma^{\mu][\rho}\chi^{\nu]}+\bar{\mpsi}^{[\rho}\gamma^{\nu][\sigma}\chi^{\mu]}\)\; .
    \end{aligned}
\label{eqn:scu_G}
\end{equation}

Starting from the linearised coupling lagrangian \eqref{eqn:new_min_min_coupling}, we can extrapolate the following nonlinear on-shell lagrangian: 
\begin{equation}
\mathcal{L}=\mathcal{L}_\text{sugra}+\mathcal{L}_\text{SFP}+\mathcal{L}_{\psi}+\mathcal{L}_\text{non-min}+\mathcal{L}_\text{quartic}+\dots\;,
 \label{full lagrangian}
\end{equation}

\noindent where the dots stand for higher order terms in the fields that are not determined in our analysis, as well as the quartic terms coming from the field redefinitions necessary to eliminate the terms proportional to the linear massive field equations. The term $\mathcal{L}_\text{sugra}$ is the standard kinetic term for the massless supergravity multiplet: 
\begin{equation}
e^{-1}\mathcal{L}_\text{sugra}=\frac{1}{2\kappa^2}R-\frac{1}{2}\bpsi_\mu\gamma^{\mu\nu\rho}D_\nu\psi_\rho  \;.
 \label{sugra lagrangian}
\end{equation}

\noindent The term $\mathcal{L}_\text{SFP}$ is the kinetic term for the massive spin-2 multiplet which arises simply as the straightforward covariantisation of the supersymmetric Fierz--Pauli lagrangian \eqref{eqn:L_SFP}:
\bea
	e^{-1}\mathcal{L}_\textup{SFP}&=&-\frac{3}{8}F^{\mu\nu}F_{\mu\nu}-\frac{3}{4}m^2A^{\mu}A_\mu -\frac{1}{2}\bar{\mpsi}_\mu\gamma^{\mu\nu\rho}D_\nu\mpsi_\rho-\frac{1}{2}\bchi_\mu\gamma^{\mu\nu\rho}D_\nu\chi_\rho-m\bar{\mpsi}_\mu\gamma^{\mu\nu}\chi_\nu\\
			&&\hspace{-2mm}+\frac{1}{4}\nabla_\mu h_{\nu\rho}\nabla^\nu h^{\mu\rho}-\frac18\nabla_\rho h_{\mu\nu}\nabla^\rho h^{\mu\nu}-\frac14\nabla_\mu h^{\mu\nu}\nabla_\nu h+\frac18\nabla_\mu h\nabla^\mu h-\frac{m^2}{8}\(h^{\mu\nu}h_{\mu\nu}-h^2\)\;. \nonumber
\label{covariant SFP lagrangian}
\eea

\noindent The term $\mathcal{L}_{\psi}$ contains all the couplings to the gravitino:
\begin{equation}
\begin{aligned}
\mathcal{L}_{\psi}=-\frac{1}{4}\varepsilon^{\mu\nu\rho\sigma}\bpsi_\mu&\biggl[\(mh_{\rho\tau}+\tilde{F}_{\rho\tau}-2i\nabla_\tau A_\rho\gamma_5\)\gamma^\tau\gamma_\nu\mpsi_\sigma \\
&\quad+\(\nabla_\lambda h_{\tau\rho}\gamma^{\lambda\tau}\gamma_5-\frac{i}{2m}\nabla_\rho F_{\lambda\tau}\gamma^{\lambda\tau}+imA_\tau\(2\delta^\tau_\rho+\gamma_\rho{}^\tau\)\)\gamma_\nu\gamma_5\chi_\sigma\biggr]\\
-\frac{ie}{4m}\bar{\rho}_{\mu\nu} \biggl[&\(mh^{\nu}{}_\lambda+\tilde{F}^{\nu}{}_\lambda-2i\nabla_\lambda A^{\nu}\gamma_5\)\gamma^{\lambda}\gamma_5\chi^{\mu}  \\
    &+\(\nabla_\rho h^{\mu}{}_\sigma\gamma^{\rho\sigma}\gamma_5-\frac{i}{2m}\nabla^{\mu}F_{\rho\sigma}\gamma^{\rho\sigma}+2imA^{\mu}-imA_\rho\gamma^{\rho\mu}\)\mpsi^{\nu} \biggr]\;,
\end{aligned}
\label{gravitino couplings lagrangian}
\end{equation}

\noindent where $\rho_{\mu\nu}=D_\mu\psi_\nu-D_\nu\psi_\mu$ is the gravitino field-strength. The term $\mathcal{L}_\text{non-min}$ includes the non-minimal couplings to gravity:
\bea
	\mathcal{L}_\text{non-min}&=&\frac{e}4 R_{\mu\sigma\nu\rho}\[-\frac{1}{2}h^{\rho\mu}h^{\sigma\nu}+\frac{1}{2}g^{\rho\mu}h^{\sigma}{}_\lambda h^{\lambda\nu}+2g^{\rho\mu}A^{\sigma}A^{\nu}-\frac{3}{4m^2}F^{\mu\sigma}F^{\nu\rho}+\frac{1}{2m^2}g^{\rho\mu}F^{\sigma}{}_\lambda F^{\lambda\nu}\right.\nonumber \\
	&&\left.+\frac{2}{m^2} g^{\sigma \nu}F^{\rho}{}_\lambda \nabla^\lambda A^{\mu}+\frac{e^{-1}}{m} \varepsilon^{\mu\sigma\lambda\tau} \nabla_\lambda h^\rho{}_\tau A^\nu +\frac{1}{m}g^{\rho\mu} \tilde{F}^{\sigma}{}_{\lambda}h^{\lambda\nu} +\frac{2}{m} \bar{\mpsi}^{\sigma}\gamma^{\mu\rho}\chi^{\nu}\]\; .
    \label{non-minimal couplings lagrangian}
\eea
Finally, the term $\mathcal{L}_\text{quartic}$ contains the quartic terms coming from integrating out the auxiliary fields $V_\mu$ and $C_{\mu\nu}$ in $- 2  V_\mu G^\mu - \frac{1}{2} V_\mu J^\mu + G^\mu X_\mu \approx - \frac{1}{4} X_\mu J^\mu$ :
\beq
\begin{aligned}
\mathcal{L}_\text{quartic}=&-\frac{1}{4}\[\varepsilon_{\mu}{}^{\nu\rho\sigma}\(\bar{\mpsi}_\rho\gamma_\nu\mpsi_\sigma-\bchi_\rho\gamma_\nu\chi_\sigma\)+\frac{4ie}{m}\nabla^\nu\(\bchi_{[\mu}\gamma_5\mpsi_{\nu]}\)\]\times\\
&\times\[-\frac{e^{-1}}{8}\varepsilon^{\mu\tau\lambda\beta}\nabla_\tau h^\alpha{}_{\lambda} h_{\beta\alpha}+\frac{9}{8}A_{\tau}\tilde{F}^{\tau\mu} +\frac{1}{4m^2}\nabla_\lambda\nabla_\beta\(\tilde{F}^{\mu\lambda}A^\beta\)\right. \\
    &\left. \quad+\frac{3}{8m}\nabla_\tau\(\nabla^\mu A^\lambda h^\tau{}_\lambda+\nabla_\lambda A^\tau h^{\mu\lambda}\)-\frac{1}{8m}\nabla_\tau\(\nabla^\lambda A^\mu h^\tau{}_\lambda+\nabla^\tau A_\lambda h^{\mu\lambda}\)\right. \\
    &\left. \quad+\frac{3m}{4}A_{\tau}h^{\tau\mu}+\frac{i}{4}\bar{\mpsi}^\tau\gamma_5 \gamma^\mu\mpsi_\tau +\frac{e^{-1}}{4m}\varepsilon^{\mu\tau\lambda\beta}\(\bchi_\tau D_\lambda\mpsi_\beta+\bar{\mpsi}_\tau D_\lambda\chi_\beta\)\]\; .
\end{aligned}   
\label{quartic lagrangian}
\eeq
Note that the non-minimal coupling of the spin-2 field in \eqref{non-minimal couplings lagrangian} could be absorbed in a redefinition of the covariant lagrangian \eqref{covariant SFP lagrangian}, by replacing $ \frac{1}{4}\nabla_\mu h_{\nu\rho}\nabla^\nu h^{\mu\rho}\rightarrow \frac{3}{8}\nabla_\mu h_{\nu\rho}\nabla^\nu h^{\mu\rho}-\frac{1}{8} \nabla^\nu h_{\nu\rho}\nabla_\mu h^{\mu\rho} $, but we prefer to define the minimal coupling from the linearisation of the Einstein--Hilbert action in a curved background. 

As we stressed, the form of the three-point coupling found in Eq.~\eqref{eqn:scu_tot}--\eqref{eqn:scu_X} is necessary to ensure a consistent coupling to off-shell new-minimal supergravity, and is unique up to four derivatives. It is relevant to ask about further generalisations. The first assumption one may relax is the property to couple consistently to new-minimal supergravity and instead couple to non-minimal supergravity. In this situation we would consider the most general supercurrent improvement \eqref{GeneralImproved}. One finds that the supersymmetry current and the stress-energy tensor get modified in the presence of $U= \mathcal{U}|_{\vartheta = 0}$ by improvement terms as 
\be T^{\mu\nu} \rightarrow T^{\mu\nu} + ( \partial^\mu \partial^\nu - \eta^{\mu\nu} \Box ) U \; , \quad  \Xi^\mu \rightarrow \Xi^\mu - 2 \gamma^{\mu\nu} \partial_\nu \Upsilon \; , \label{eqn:U_imp_T_Xi}  \ee
where 
\be \delta U  = \bar \epsilon \Upsilon \; . \ee
In principle we may consider $U$ as a general polynomial in the fields and their derivatives including up to two derivatives, as we did for the new-minimal coupling. However, the modification of the three-point coupling then reads 
\be \mathcal{L}\rightarrow \mathcal{L} - \frac{e}{2} R\,  U +e \bar \rho_{\mu\nu} \gamma^{\mu\nu} \Upsilon \; , \label{eqn:non_min_imp_lag} \ee
and they  can be eliminated at three-point by a field redefinition of the massless fields. We can therefore choose to disregard them at three-point. In fact this is also true for any term which couples non-minimally to the Ricci tensor in \eqref{non-minimal couplings lagrangian}. For the sake of completeness we can write the full non-linear lagrangian which cannot be eliminated at three points
\bea
	e^{-1}\mathcal{L}&=&\frac{1}{2\kappa^2}R-\frac{1}{2}\bpsi_\mu\gamma^{\mu\nu\rho}D_\nu\psi_\rho  \label{on-shell non zero lagrangian} \\
	&&-\frac{3}{8}F^{\mu\nu}F_{\mu\nu}-\frac{3}{4}m^2A^{\mu}A_\mu -\frac{1}{2}\bar{\mpsi}_\mu\gamma^{\mu\nu\rho}D_\nu\mpsi_\rho-\frac{1}{2}\bchi_\mu\gamma^{\mu\nu\rho}D_\nu\chi_\rho-m\bar{\mpsi}_\mu\gamma^{\mu\nu}\chi_\nu\nonumber \\
			&&+\frac{1}{4}\nabla_\mu h_{\nu\rho}\nabla^\nu h^{\mu\rho}-\frac18\nabla_\rho h_{\mu\nu}\nabla^\rho h^{\mu\nu}-\frac14\nabla_\mu h^{\mu\nu}\nabla_\nu h+\frac18\nabla_\mu h\nabla^\mu h-\frac{m^2}{8}\(h^{\mu\nu}h_{\mu\nu}-h^2\) \nonumber\\
			&&-\frac{e^{-1}}{4 }\varepsilon^{\mu\nu\rho\sigma}\bpsi_\mu\biggl[\(mh_{\rho\tau}+\tilde{F}_{\rho\tau}-2i\nabla_\tau A_\rho\gamma_5\)\gamma^\tau\gamma_\nu\mpsi_\sigma\nonumber \\
&&\quad+\(\nabla_\lambda h_{\tau\rho}\gamma^{\lambda\tau}\gamma_5-\frac{i}{2m}\nabla_\rho F_{\lambda\tau}\gamma^{\lambda\tau}+imA_\tau\(2\delta^\tau_\rho+\gamma_\rho{}^\tau\)\)\gamma_\nu\gamma_5\chi_\sigma\biggr]\nonumber\\
&&-\frac{i}{4m}\bar{\rho}_{\mu\nu} \biggl[\(mh^{\nu}{}_\lambda+\tilde{F}^{\nu}{}_\lambda-2i\nabla_\lambda A^{\nu}\gamma_5\)\gamma^{\lambda}\gamma_5\chi^{\mu} \nonumber \\
   & &\hspace{20mm} +\(\nabla_\rho h^{\mu}{}_\sigma\gamma^{\rho\sigma}\gamma_5-\frac{i}{2m}\nabla^{\mu}F_{\rho\sigma}\gamma^{\rho\sigma}+imA^{\mu}\)\mpsi^{\nu} \biggr]\nonumber\\
	&&+\frac14 R_{\mu\sigma\nu\rho}\[-\frac{1}{2}h^{\rho\mu}h^{\sigma\nu}-\frac{3}{4m^2}F^{\mu\sigma}F^{\nu\rho}+\frac{e^{-1}}{m} \varepsilon^{\mu\sigma\lambda\tau} \nabla_\lambda h^\rho{}_\tau A^\nu +\frac{2}{m} \bar{\mpsi}^{\sigma}\gamma^{\mu\rho}\chi^{\nu}\]+\dots\; ,\nonumber
  \eea
where the dots include higher order terms that we have not computed as well as quartic terms in the massive fields that come from the field redefinition necessary to eliminate the terms linear in the Ricci tensor and Rarita--Schwinger field equation and quartic terms quadratic in the massless fields necessary to eliminate the terms proportional to the linear massive field equations. Writing $g_{\mu\nu} = \eta_{\mu\nu} + 2 \kappa g_{\mu\nu}^{\scalebox{0.6}{(1)}}$, we can write the on-shell three-point coupling of two massive spin-2 fields and a massless graviton as
\begin{multline} \frac{\kappa}{2} \Bigl(  g_{\mu\nu}^{\scalebox{0.6}{(1)}} \bigl( \partial^\mu h^{\rho\sigma} \partial^\nu h_{\rho\sigma} - 4 \partial^\mu h^{\rho\sigma} \partial_\rho h^{\nu}{}_\sigma \bigr) + h^{\mu\nu} \bigl( \partial_\mu g_{\rho\sigma}^{\scalebox{0.6}{(1)}}  \partial_\nu h^{\rho\sigma} -\partial_\rho g_{\mu\sigma}^{\scalebox{0.6}{(1)}}  \partial_\nu h^{\rho\sigma} \bigr) \\+ \partial_\mu \partial^\rho g_{\nu\rho}^{\scalebox{0.6}{(1)}}  \bigl(2 h^{\mu}{}_\sigma h^{\nu\sigma}- \tfrac14 \eta^{\mu\nu} h_{\sigma\lambda} h^{\sigma\lambda} \bigr)\Bigr) \;  , \end{multline}
which reproduces the coupling of a massive open string state of mass $m = \frac{1}{\sqrt{\alpha'}}$ to the graviton in superstring theory \cite{Lust:2021jps} (and in bosonic string theory \cite{Buchbinder:1999ar}). The massive open string theory states also have a self-interaction at three-point with leading component in $m^2 \kappa h_{\mu\nu} h^{\mu\rho} h^{\nu}{}_\rho$ \cite{Lust:2021jps}, which can be supersymmetrised using the full-superspace integral of a superfield $S$ with first component $S|_{\vartheta=0}=\kappa h^{\mu\nu} A_\mu A_\nu$. In our analysis these terms are not required by supersymmetry and can thus appear with arbitrary overall coefficients. 

We find that four-derivative terms are necessary and unique assuming a consistent coupling to new-minimal supergravity, but in principle one may consider higher derivative deformations of the 
 improvement \eqref{eqn:Dj_fin}. This analysis is rather non-trivial already at the 6-derivative level and we leave it for future work. Even more generally one may consider that the spin-2 supermultiplet does not couple consistently to non-minimal off-shell supergravity. This is possible if the couplings modify the supersymmetry algebra and we will explain in Section \ref{sec:stuckelberg} that this is in particular the case for the Kaluza--Klein theory obtained by reduction on $S^1 / \mathbb{Z}_2$.


\section{The 2 \texorpdfstring{$\to$}{->} 2 scattering amplitude of the massive spin-2 field} \label{sec:amp}

In the previous sections we determined the consistent couplings of the massive spin-2 supermultiplet of Eq.~\eqref{eqn:ms2_susy} to supergravity at leading order in $M_{\scalebox{0.6}{P}}$. As we emphasised, when restricting to the new-minimal off-shell formulation of supergravity \eqref{eqn:nm_susy} \cite{Sohnius:1981tp}, this coupling is unique at the 4-derivative order and it is determined by the supercurrent supermultiplet $\S^\mu$ of Eq.~\eqref{eqn:scu_tot}--\eqref{eqn:scu_X}. In the non-minimal off-shell formulation of supergravity  \cite{Siegel:1978mj} one can consider a more general current supermultiplet including a coupling of the type displayed in Eq.~\eqref{eqn:non_min_imp_lag}.

In this section we analyse the $2\to2$ tree-level elastic scattering amplitude of the massive spin-2 field $h_{\mu\nu}$ in this theory. The only contribution to this amplitude comes from the full  stress-energy tensor \eqref{eqn:scu_T} combining  the Fierz--Pauli stress-energy tensor $T^{\mu\nu}_\textup{FP}$ given in Eq.~\eqref{eqn:sc_T_K} and the non-minimal coupling \eqref{eqn:scu_G}:
\beq\begin{aligned} \includegraphics[scale=0.5]{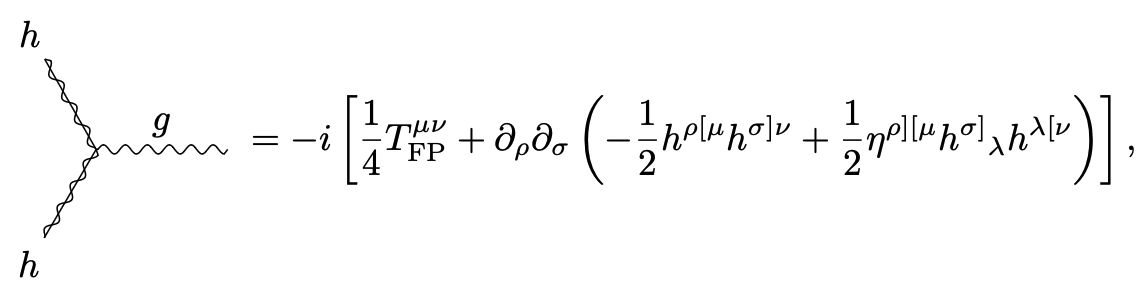} \end{aligned}\label{three point function}\eeq

We compute the amplitude in the case where cubic and quartic couplings in the massive fields are not included. We have not analysed them in Sections  \ref{sec:sc} and \ref{sec:ms2_sugra}, although they can in principle be added consistently with supersymmetry. Note that one could eliminate the terms linear in the Ricci tensor in \eqref{non-minimal couplings lagrangian} by a field redefinition at the price of introducing quartic couplings in the massive fields. Here we work instead with the complete three-point couplings and no quartic term. There are also four-point couplings quadratic in the massless fields which are required by non-linear supersymmetry. However, being quadratic in the massless fields they do not contribute at tree-level to the massive spin-2 two-to-two amplitude. Therefore, we consider only the contributions of the three-point function \eqref{three point function} to the tree-level massive spin-2 two-to-two amplitude, which amounts to the sum of the $s$, $t$ and $u$-channel graviton interaction.

\beq \begin{aligned}\includegraphics[scale=0.6]{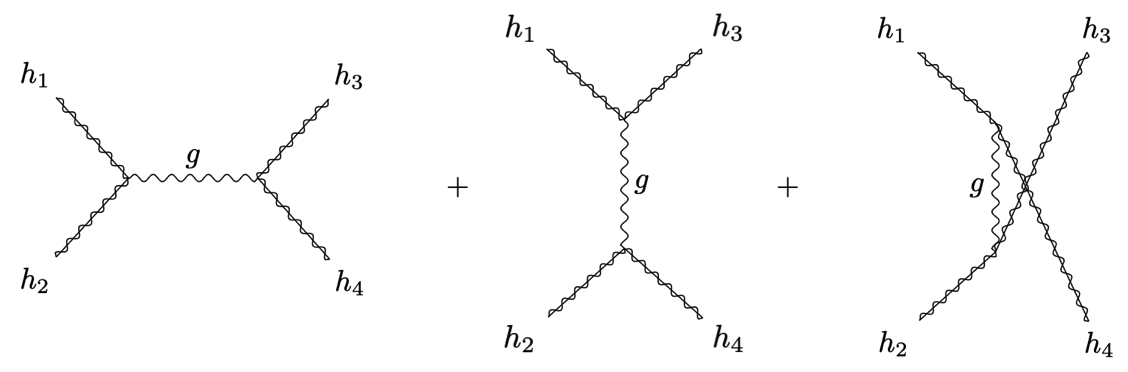}\end{aligned}\eeq

\noindent Each diagram appears as one contraction of the free-field creation and annihilation operators in 
\beq
  i  (2\pi)^4 \delta^{\scalebox{0.6}{(4)}}(p_1+p_2-p_1'-p_2') \A=-\frac{\kappa^2}{2}  \langle {\rm out} |  \int \hspace{-2mm} d^4x d^4 y :T^{\mu\nu}_{h}(x)D_{\mu\nu\rho\sigma}(x-y)T^{\rho\sigma}_{h}(y):\;  |{\rm in}\rangle  \; ,
\label{eqn:amp_ch}
\eeq
where  $D_{\mu\nu\rho\sigma}$ is the graviton propagator, which we define in momentum space as 
\beq
    D_{\mu\nu\rho\sigma}(p)=-\frac{i}{2p^2}\(\eta_{\mu\rho}\eta_{\nu\sigma}+\eta_{\mu\sigma}\eta_{\nu\rho}-\eta_{\mu\nu}\eta_{\rho\sigma}\) \ .
\eeq

The expansion of the massive spin-2 field in momentum space is 
\beq
    h_{\mu\nu}(x)=2 \int \frac{d^3p}{\(2\pi\)^{\frac32}\sqrt{2E_{{\boldsymbol p}}}}\sum_{\sigma}\(\epsilon_{\mu\nu}\(\sigma,\boldsymbol{p}\)e^{ip\cdot x}a_{\sigma,\boldsymbol{p}}+\bepsilon_{\mu\nu}\(\sigma,\boldsymbol{p}\)e^{-ip\cdot x}a_{\sigma,\boldsymbol{p}}^\dagger\)\ ,
\eeq
where the factor of $2$ is due to the non-canonical normalisation of $h_{\mu\nu}$ in Eq.~\eqref{covariant SFP lagrangian}. We write $\{a_{\sigma,\boldsymbol{p}},a_{\sigma,\boldsymbol{p}}^\dagger\}$ the ladder operators and $\epsilon_{\mu\nu}(\sigma,\boldsymbol{p})$ the polarisation tensors, both depending on the space momentum $\boldsymbol p$ and on the five polarisation states of the massive spin-2 field labeled by the helicity $\sigma=\left\{\pm2,\pm1,0\right\}$. These polarisation tensors are such that
\begin{align}
    p^\mu\epsilon_{\mu\nu}(\sigma,\bp)=0\; , \qquad  \epsilon^\mu{}_\mu(\sigma,\bp)=0\; ,
\end{align}

\noindent consistently with the equations of motion \eqref{eqn:eom_full}-\eqref{eqn:eom}, and satisfy the following orthogonality and completeness relations:
\beq
\begin{aligned}
    \epsilon^{\mu\nu}(\sigma,\bp)\bepsilon_{\mu\nu}(\sigma',\bp)&=\delta_{\sigma\sigma'},\\
    \sum_\sigma\epsilon^{\mu\nu}(\sigma,\bp)\bepsilon^{\rho\sigma}(\sigma,\bp)&=\frac{1}{2}\(P^{\mu\rho}P^{\nu\sigma}+P^{\nu\rho}P^{\nu\sigma}\)-\frac{1}{3}P^{\mu\nu}P^{\rho\sigma},
\end{aligned}
\eeq

\noindent with $\textstyle{P^{\mu\nu}=\eta^{\mu\nu}+\frac{p^\mu p^\nu}{m^2}}$. Parameterising the momentum $p^{\mu}=\(E_{\boldsymbol{p}},\boldsymbol{p}\)$ as
\beq
\begin{aligned}
	 E_{\boldsymbol{p}}=\sqrt{\boldsymbol{p}^2+m^2 }, &&&& \boldsymbol{p}=\abs{\boldsymbol{p}}\(\sin\theta\cos\phi,\sin\theta\sin\phi,\cos\theta\)\equiv\abs{\boldsymbol{p}}\hat{\bp},
\end{aligned}
\eeq

\noindent with $\(\theta,\phi\)$ being the angles with respect to the $z$-axis, we can choose the spin-2 polarisation tensors  to be \cite{SekharChivukula:2019yul,Chivukula:2020hvi}
\beq
\begin{aligned}
    \epsilon^{\mu\nu}\({\pm2},\bp\)&=\epsilon^\mu_{\pm 1}(\bp)\epsilon^\nu_{\pm 1}(\bp), \\
    \epsilon^{\mu\nu}\(\pm1,\bp\)&=\frac{1}{\sqrt{2}}\(\epsilon^\mu_{\pm1}(\bp)\epsilon^\nu_{0}(\bp)+\epsilon^\nu_{\pm1}(\bp)\epsilon^\mu_{0}(\bp)\), \\
    \epsilon^{\mu\nu}(0,\bp)&=\frac{1}{\sqrt{6}}\(\epsilon^\mu_{1}(\bp)\epsilon^\nu_{-1}(\bp)+\epsilon^\nu_{1}(\bp)\epsilon^\mu_{-1}(\bp)+2\epsilon^\mu_{0}(\bp)\epsilon^\nu_{0}(\bp)\),
\end{aligned}
\eeq

\noindent with
\beq
\begin{aligned}
    \epsilon_{\pm1}^\mu(\bp)=&\pm\frac{e^{\pm i\phi}}{\sqrt{2}}\(0,-\cos\theta\cos\phi\pm i\sin\phi,-\cos\theta\sin\phi\mp i\cos\phi,\sin\theta\)\; , \\
    \epsilon_0^\mu(\bp)=&\frac{E_{\bp}}{m}\( \frac{|{\bf p}|}{E_{\bp}},\hat{\bp}\)\; .
\end{aligned}
\eeq

With these tools, we can compute the scattering amplitude \eqref{eqn:amp_ch} as described above. In the absence of the improvement term due to supersymmetry, it is the amplitude with all helicities  $\sigma_a=0$ that scales with the highest power of $s$  \cite{SekharChivukula:2019yul,Chivukula:2020hvi}. This is because the associated polarisation tensor becomes approximately longitudinal at higher energy. With the improvement term~\eqref{eqn:sc_T_K} one computes instead that 
\begin{multline}
  \A(0,0,0,0) = \frac{\kappa^2}{16 m^4 s (s - 4 m^2)} \Biggl( 
  s^5 \sin^2\theta \bigl(3 + \cos^2\theta\bigr) - \frac{4 m^2s^4}{3}  (21 - 29 \cos^2\theta) \\+ 
   2 m^4 s^3 \biggl( \frac{ 32}{\sin^2\theta} + \frac{37 - 231  \cos^2\theta + 18  \cos^4\theta}{3}  \biggr) - 
   16 m^6 s^2 \biggl(\frac{ 16}{\sin^2\theta} - 7 + 2  \cos^2\theta  - \cos^4\theta\biggr) \\+ 
   32 m^8 s\,  \biggl(\frac{ 4}{\sin^2\theta} + \frac{11 + 26  \cos^2\theta}{3}  \biggr) +128 m^{10}\cos(2\theta) \Biggr)
\end{multline}
is well behaved in the Regge limit, \textit{i.e.} the amplitude grows as $\A(0,0,0,0)\sim s^2$ for  $s\gg -t$ and $s\gg m^2$. At fixed angle $\theta$, it grows as $\A(0,0,0,0)\sim s^3$, which is the softest possible behaviour for a massive spin-2 scattering amplitude according to  \cite{Arkani-Hamed:2002bjr,Arkani-Hamed:2003roe,Schwartz:2003vj,Bonifacio:2018vzv,Bonifacio:2018aon,Bonifacio:2019mgk,Kundu:2023cof}. Using the notation \eqref{eqn:cutoff}, this would naively lead to a cutoff $\Lambda_3$. However, the higher-energy behaviour of the amplitude is not improved for all helicities, in particular when two of the helicities are equal to $\pm 1$. For instance, the amplitude 
\begin{multline}
  \A(1,1,0,0) = \frac{\kappa^2}{384 m^6 s (s - 4 m^2)} \Biggl( s^6 \bigl(5 - \cos^2\theta\bigr)- m^2 s^5 \bigl(49 - 9 \cos^4\theta\bigr) \\ +2 m^4 s^4 \bigl(217 - 113 \cos^2\theta + 26\cos^4\theta\bigr) - 
  4 m^6 s^3 \bigl(480 - 265  \cos^2\theta + 69 \cos^4\theta \bigr)\\[3mm]+  32 m^8 s^2 \bigl(103 + 96 \cos^2\theta - 9 \cos^4\theta \bigr)\\ - 
  64 m^{10} s \bigl(56 + 119 \cos^2\theta + \cos^4\theta\bigr) 
- 3072 m^{12}\cos(2\theta) \Biggr)
\end{multline}
grows as $ \A(1,1,0,0)\sim s^4$ in both the Regge and the large energy limit at fixed angle. The amplitudes  $\A(0,0,\pm 1,\pm 1)$, $\A(\pm 1,\pm 1,0,0)$, $\A(0,\pm 1,\pm 1,0)$, $\A(\pm 1,0,0,\pm 1)$, as well as  $ \A(\pm 1,0,\mp 1 ,0)$ and $\A(0,\pm 1,0,\mp 1 )$ scale in the same way as $s^4$ in both the Regge limit and the  large energy limit at fixed angle. 

Therefore, the coupling to new-minimal supergravity of the massive spin-2 multiplet \eqref{eqn:ms2_susy} described by the supercurrent $\S^\mu$ of Eq.~\eqref{eqn:scu_tot} is associated to the cutoff $\Lambda_4$, according to Eq.~\eqref{eqn:cutoff}. Supersymmetry famously improves the behaviour of the scattering amplitudes and in this case it tunes the interactions between the massive spin-2 field and gravity in such a way that the associated growth in energy is slightly better than the common $\Lambda_5$ case. However, one would need to include self-interactions to get to the $\Lambda_3$ cutoff  \cite{Bonifacio:2018aon}. 

We now turn to the couplings to non-minimal off-shell supergravity \eqref{GeneralImproved}-\eqref{eqn:non_min_imp_lag}. In this formulation the stress-energy tensor receives additional  contributions of the type~\eqref{eqn:U_imp_T_Xi}, proportional to the composite operator $U= \mathcal{U}|_{\vartheta = 0}$ of Eq.~\eqref{GeneralImproved}, which is quadratic in fields and contains up to two derivatives. Hence, the possible nontrivial contributions to the $2\to2$ scattering amplitudes of the massive spin-2 field $h_{\mu\nu}$ come from the stress-energy tensor 
\beq
	T^{\mu\nu}_{U}=\(\partial^\mu\partial^\nu-\eta^{\mu\nu}\Box\)\(c_1 h^{\rho\sigma}h_{\rho\sigma}+c_2 \partial_\lambda  h^{\rho\sigma}\partial^\lambda h_{\rho\sigma} + c_3 \partial_\rho \partial_\sigma ( h^{\rho\lambda} h^{\sigma}{}_\lambda ) \) \ .
\label{eqn:U_terms_amp}
\eeq

\noindent One computes that adding these terms to the stress-energy tensor results in a $2\to2$ scattering amplitude of the massive spin-2 field that grows as $\A(0,0,0,0)\sim s^5$ or more in the large $s$ limit. 

Hence, the theory defined by the supercurrent superfield $\mathcal{S}^\mu$ in Eq.~\eqref{eqn:scu_tot} is not only the unique coupling to new-minimal off-shell supergravity but it is also the only coupling to supergravity with tree-level cutoff scale $\Lambda_4$ instead of $\Lambda_5$.  All the other couplings of the spin-2 field to gravity that one can obtain by moving to a non-minimal off-shell formulation through $\mathcal{U}$-like improvements to the supercurrent \eqref{GeneralImproved} are associated with a lower cutoff scale $\Lambda_5$.\footnote{Note however that one can get the same cutoff scale $\Lambda_4$ for a one-parameter family of improvement terms if one does not require supersymmetry.}

%

\section{The St\"uckelberg symmetry in Kaluza--Klein supergravity} \label{sec:stuckelberg}
In this paper we have found that there is a unique coupling of off-shell new-minimal supergravity to a massive spin-2 multiplet in the unitary gauge with at most four-derivative interactions. This coupling involves in particular four-derivative terms, although we have seen in the preceding section that the $2\rightarrow2$ scattering amplitude for massive spin-2 particles has the best possible behaviour at high energy. This may seem in contradiction with the coupling expected from the Kaluza--Klein reduction of minimal supergravity in five dimensions, which yields indeed a two-derivative theory. In this section we wish to clarify this point. 

The assumption that the massive multiplet is coupled to off-shell new-minimal supergravity is non-trivial. Starting from the minimal off-shell formulation of pure supergravity in five dimensions \cite{Zucker:1999ej}, the $\mathbb{Z}_2$ orbifold projects the massless sector to a non-minimal off-shell formulation  of supergravity \cite{Siegel:1978mj,Zucker:2000ks}. The non-minimal supergravity multiplet is reducible and includes a 16+16 irreducible component \cite{Girardi:1984vq}  dual to the general energy-momentum tensor supermultiplet \cite{Komargodski:2010rb}. It must be possible to write the orbifold theory on $S^1 / \mathbb{Z}_2$ in the non-minimal off-shell formulation of supergravity, but this formulation is by construction written in terms of St\"uckelberg fields with the non-linear St\"uckelberg gauge invariance coming from five-dimensional gauge invariance. On the contrary, the analysis we have carried out in this paper is for the St\"uckelberg field strengths, or equivalently, in the unitary gauge of the St\"uckelberg symmetry. We shall argue that  one cannot write the orbifold theory in the unitary gauge for the massive fields using non-minimal supergravity. 

Let us consider the Kaluza--Klein ansatz for the five-dimensional vielbeins $E^A$, the vector field $\A$ and the gravitino potential $\Psi$
\begin{align}
E^a&=\phi^{-\frac{1}{2}}e^a\; , \qquad 
E^4=\phi\(dy+B\)\; ,  \qquad \A=a\(dy+B\)+A\; , \nonumber \\
\Psi&=\phi^{\frac{5}{4}}\,\zeta\(dy+B\)+\phi^{-\frac{1}{4}}\(\psi-\frac{1}{2}e^a\gamma_a\gamma_5\zeta\)\; , \label{KK_ansatz} 
\end{align}
where $e_\mu{}^a$, $\phi$ and $a$ are even functions of the circle coordinate $y$, the bosonic fields $A_\mu$ and $B_\mu$ are odd and the fermionic ones are such that 
\beq
\psi{}_\mu(-y)=\sigma_3\gamma_5\psi{}_\mu(y)\; , \qquad  \zeta(-y)=-\sigma_3\gamma_5\zeta(y)\; ,
\eeq
with $\sigma_3$ the $R$-symmetry isospin Pauli matrix.
We refer to Appendix \ref{app:kk} for more details on the Kaluza--Klein ansatz. In the unitary gauge, one sets 
\be B_\mu(x,y) = 0 \; , \quad \phi(x,y) = \phi(x) \; , \quad a(x,y) = a(x) \; , \quad \zeta(x,y) = \tfrac12\bigl( 1-\sigma_3\gamma_5\bigr) \zeta(x)\; .  \ee 
By construction, the lagrangian obtained from five dimensions with this ansatz only includes terms with at most two derivatives. However, the supersymmetry transformations must be modified to preserve the gauge. Including the five-dimensional diffeomorphism parameter $\xi^M(x,y)$, one obtains for example 
\begin{align} \delta B_\mu(x,y) &=\frac{\phi(x)^{-\frac{3}{2}}}{2}\bar{\epsilon}(x)  \gamma_5\psi_\mu(x,y)   + \phi(x)^{-3} g_{\mu\nu}(x,y) \partial_y \xi^\nu(x,y) + \partial_\mu \xi^5(x,y)  \nonumber\\
\delta \phi(x,y) &= \frac{\phi(x)}{2} \bar \epsilon(x) \gamma_5 \zeta(x) + \xi^\mu(x,y) \partial_\mu \phi(x) + \partial_y \xi^5(x,y) \phi(x) \; , \end{align}
where one must set 
\begin{align}
 \partial_y \xi^5(x,y) &= - \xi^\mu(x,y) \partial_\mu \log \phi(x)\; , \nonumber \\
  \partial_y \xi^\mu(x,y) + \phi(x)^3 g^{\mu\nu}(x,y) \partial_\nu \xi^5(x,y) &= - \frac{\phi(x)^{\frac{3}{2}}}{2} g^{\mu\nu}(x,y) \bar{\epsilon}(x)\gamma_5\psi_\nu(x,y)\end{align}
in order to preserve the unitary gauge for the spin-2 St\"uckelberg fields. One must similarly solve for the mode expansion of the spinor parameter $\epsilon(x,y)$ to preserve the  unitary gauge for the spin-$\frac32$ St\"uckelberg fields, but for brevity we shall focus on the bosons in this discussion. Preserving the unitary gauge requires to modify the supersymmetry transformations of the vielbeins accordingly: 
\beq \delta e_\mu{}^a(x,y) = \frac{1}{2}\bar{\epsilon}(x)\gamma^a\psi_\mu(x,y) + D_\mu \bigl(  \xi^\nu(x,y)e_\nu{}^a(x,y) \bigr) + \xi^5(x,y) \partial_y e_\mu{}^a(x,y)\; . \eeq
In the linearised approximation the second term is responsible for the $\frac1{m}$ term in the supersymmetry variation of the massive spin-2 field, but the supersymmetry transformation of the (massless) four-dimensional vielbeins is also modified by quadratic terms in the massive fields involving an additional derivative. Moreover, the supersymmetry algebra itself is modified since the circle diffeomorphism algebra implies that the commutator of two supersymmetries now produces an additional diffeomorphism with a field dependent parameter quartic in the massive fields 
\beq \xi^\mu(\epsilon_1,\epsilon_2) \sim - \frac12  \bar \epsilon_1 \gamma^\mu \epsilon_2 + \frac{1}{4m^2}  \bar \epsilon_1 h^{\nu\rho} \psi_\rho  \partial_\nu \bigl(  \bar \epsilon_2 h^{\mu\sigma} \psi_\sigma \bigr) - \frac{1}{4m^2}  \bar \epsilon_2 h^{\nu\rho} \psi_\rho  \partial_\nu \bigl(  \bar \epsilon_1 h^{\mu\sigma} \psi_\sigma \bigr)  \; . \eeq
This implies in particular that one cannot recover the canonical variation of the four-dimensional vielbeins through a local field redefinition in the unitary gauge. Introducing the mode of the five-dimensional supersymmetry parameter necessary to preserve the unitary gauge for the spin-$\frac32$ St\"uckelberg fields also introduces extra terms that can however be absorbed in a redefinition of the spinor parameter. This is therefore a property of the spin-2 field only. 

In order to describe the orbifold theory one needs therefore to either use all the St\"uckelberg fields and do not assume the stress-energy   tensor supermultiplets components only depend on the St\"uckelberg field strengths, or, relax the condition that the  stress-energy   tensor multiplet is a rigid supersymmetry multiplet and compute the cubic coupling assuming that the supersymmetry transformations of both the massive and the massless fields can be modified at this order. We shall not carry out this general analysis. Note nonetheless that the existence of such a deformation of the supersymmetry algebra is very non-trivial, and it is reasonable to believe that the unique consistent deformation  is the one obtained from Kaluza--Klein reduction.

\section{Discussion and conclusions} \label{sec:conclusions}

In this work we have constructed and analysed the supercurrent superfield associated to the massive spin-2 multiplet \eqref{eqn:ms2_susy}, starting from the associated $R$-symmetry current \eqref{eqn:axial_c}. Following the transformation rules \eqref{eqn:sc_susy} \cite{Sohnius:1981tp}, we determined the $R$-multiplet of currents given in Eq.~\eqref{eqn:sc_0}. This includes in particular the stress-energy tensor \eqref{eqn:sc_T_1}, which is found to contain improvement terms \eqref{eqn:sc_B}-\eqref{eqn:sc_G} with respect to the standard Hilbert stress-energy tensor \eqref{eqn:sc_T_K}, some of which being higher-derivative operators.

We analyse the conditions under which the supercurrent superfield defines a consistent coupling of the massive spin-2 multiplet \eqref{eqn:ms2_susy} to new-minimal $\mathcal{N}=1$ supergravity in four dimensions~\eqref{eqn:new_min_min_coupling}-\eqref{eqn:nm_susy}. The improvement term to the Hilbert stress-energy-momentum tensor must be associated to couplings to the Riemann tensor and {\it not} to the naked Levi--Civita connection as in~\eqref{eqn:sc_B} for consistency with non-linear diffeomorphism invariance. Assuming a coupling to  new-minimal supergravity with at most four-derivative couplings, we find that there is a unique consistent solution~\eqref{eqn:scu_coeff_sol}. We show that assuming a consistent coupling to non-minimal supergravity allows for more couplings, but that can all be reabsorbed at leading order  into a field redefinition of the vielbeins and gravitino fields. Up to this redefinition, we have therefore demonstrated that there is a unique coupling of the massive spin-2 multiplet in the unitary gauge to (undeformed) off-shell supergravity. This constitutes the main result of the paper. 

In Section \ref{sec:amp} we compute the $2\to2$ elastic scattering amplitude of the massive spin-2 field mediated by the graviton in the resulting theory. We find that the new-minimal amplitude grows as $s^4$ in the large energy limit at fixed scattering angle, while it grows as $s^5$ for a general  coupling to non-minimal supergravity. This singles out the unique coupling to new-minimal supergravity.

As we explain in Section \ref{sec:stuckelberg} and Appendix \ref{app:kk}, this does not include the case of a massive spin-2 multiplet obtained from five-dimensional pure supergravity on $S^1/\mathbb{Z}_2$, with a $\mathbb{Z}_2$ orbifold action breaking $\mathcal{N}=2$ to $\mathcal{N}=1$ supersymmetry. The reason is that, although supergravity in five dimensions can be defined off-shell consistently with the $\mathbb{Z}_2$ orbifold action, the massive fields must be realised \`a la St\"{u}ckelberg in order to preserve the algebra of local supersymmetry. Assuming instead that the massive spin-2 multiplet is realised in the unitary gauge, we showed that the algebra of supersymmetry is modified by higher-derivative terms and is inconsistent with the off-shell formulations of supergravity in four dimensions. It is known that there is no non-trivial deformation of the diffeormorphism algebra \cite{Barnich:1995ap}, but Kaluza--Klein theory including massive states in the unitary gauge provides an example of a non-trivial deformation of the local supersymmetry algebra. In this case the deformation descends from a subalgebra of the local supersymmetry algebra in five dimensions. Although we have not carried out the general analysis, it is reasonable to believe that this deformation is unique, such that there would exist only two consistent ways to couple a massive spin-2 multiplet to supergravity, the one introduced in this paper and the one obtained by Kaluza--Klein reduction.


Our result has a natural interpretation in superstring theory. A non-BPS massive multiplet arises in type I superstring theory on $T^6$ from the first oscillator mode of the open string with mass $m= \frac{1}{\sqrt{\alpha'}}$ \cite{Lust:2021jps}. In $\mathcal{N}=4$ supergravity, it is a long multiplet comprising 128+128 degrees of freedom descending from a symmetric tensor $h_{MN}$, a 3-form $A_{MNP}$ and a vector spinor $\lambda_M$ in ten dimensions.  The three-point amplitude of two such spin-2 particles and a graviton gives precisely the non-minimal coupling that we obtained in this paper as the most general coupling to off-shell supergravity in four dimensions. On the other hand, the massless closed string states in ten dimensions give for each Kaluza--Klein mode along $T^6$ a massive 1/2-BPS spin-2 multiplet comprising 24+24 degrees of freedom and five massive 1/2-BPS vector multiplets. One will obtain the same amplitudes at tree-level, for either spin-2 multiplets,  if one considers an orbifold of the theory breaking $\mathcal{N}=4$ to $\mathcal{N}\ge 1$ supersymmetry. We find therefore that the coupling we defined in this paper has a natural interpretation in superstring theory as the coupling to gravity of a non-trivial spin-2 Regge excitation of the open string.\footnote{It would be interesting to also compare our result with the analogous coupling that arises in the closed string case. The gravitational coupling of a massive spin-2 closed string oscillator mode was computed in \cite{Lust:2023sfk}. However, this massive spin-2 mode does not belong to a massive spin-2 multiplet, but is instead the superposition of four spin-2 fields, respectively in a spin-2, two spin-$\frac{3}{2}$ and a spin-3 multiplet. In this case the coupling to gravity is minimal, in apparent contradiction with our result, but one would need to compute the actual spin-2 multiplet amplitude in order to make a proper comparison.} As such, we know that there exist ultra-violet completions of the theory, although string theory will always include more higher-spin particles of mass of the same order $\sqrt{\frac{n}{\alpha'}}$ for $n\in \mathbb{N}$ so that $\Lambda_3$ is far beyond the scale of new physics for $m\ll M_{\scalebox{0.6}{P}}$. 

It is therefore tempting to extrapolate our results to extended supergravity with $\mathcal{N}\ge 2$ supersymmetry. Our findings show that there is a unique way to write the gravitational coupling of a long spin-2 multiplet, with the particular non-minimal coupling including higher derivative terms for the spin-1 fields. We expect also that the gravitational coupling of a 1/2-BPS short multiplet must necessarily be the one of a Kaluza--Klein multiplet by supersymmetry, corresponding to the only non-trivial modification of the local supersymmetry algebra descending from higher dimensions. This corroborates the expectations one may get within the infinite distance swampland paradigm \cite{Ooguri:2006in,Grimm:2018ohb,Lee:2019wij}. Considering an asymptotic point in the moduli space of vacua in which the 1/2-BPS mass of such a spin-2 multiplet vanishes, we must get an infinite tower of massless spin-2 multiplets that will be interpreted as a decompactification limit. On the contrary, an asymptotic point in moduli space in which a long multiplet mass vanishes will give rise to an infinite tower of higher-spin fields becoming massless, as in the tensionless limit of a perturbative string theory.

It would be very interesting to sharpen the spin-2 conjecture proposed in \cite{Klaewer:2018yxi,Kundu:2023cof} for supersymmetric theories. The comparaison with superstring theory suggests that a quantum supergravity theory with a massive spin-2 multiplet of mass $m$ must also include an infinite Regge trajectory of massive multiplets of spin $s$ and mass $\sqrt{s-1}\,  m$ for all $s\ge 2$. In this paper we have shown that the supersymmetric coupling of the massive spin-2 multiplet brings the non-minimal coupling~\eqref{eqn:scu_G} that includes the higher derivative term $R_{\mu\nu\rho\sigma}F^{\mu\nu}F^{\rho\sigma}$ for the massive spin-1 field. It was shown  in  \cite{Hollowood:2015elj} that the propagation of a massless gauge-field in the background of a gravitational shock wave  violates causality in four dimensions. We expect a similar problem to occur for a massive vector field when $m\ll M_{\scalebox{0.6}{P}}$. It would be interesting to generalise the analysis of  \cite{Camanho:2014apa,Hollowood:2015elj} to find if the restoration of causality requires the presence of an infinite tower of higher-spin fields as for the $R^3$ correction to Einstein--Hilbert. Note, however, that the analysis of \cite{Hollowood:2015elj,Bellazzini:2021shn} applies to quantum electrodynamics in which \cite{Drummond:1979pp} finds the effective coupling $\frac{\alpha}{360 \pi m_{\rm e}^2} R_{\mu\nu\rho\sigma}F^{\mu\nu}F^{\rho\sigma}$. In this case the higher spin resonances can be interpreted as unstable positronium bound states. One should therefore be careful before interpreting a Regge trajectory of higher spin massive fields as being necessarily associated to a string spectrum. Despite these cautions, our results may suggest that string theory and Kaluza--Klein theories are the only frameworks in which the tree-level amplitudes of a massive spin-2 multiplet coupled to supergravity are consistent with unitarity and causality.

There are several ways to extend the present work. We have proved that there is a unique three-point coupling of a massive spin-2 consistent with off-shell supergravity. It would be interesting to extend this result at the non-linear level with a fully supersymmetric lagrangian such that we could study the fate of propagating ghosts in the model. We know that a non-linear theory must make sense with additional fields because it is realised in superstring theory, but it remains to check if one could find violations of causality that would require  the introduction of higher-spin multiplets with masses below the naive cut-off scale $\Lambda_4$. Also it would be interesting to study the uniqueness of the coupling when including higher derivative corrections. Another subject of improvement is the study of the supersymmetric couplings involving a deformation of the supersymmetry algebra to prove (or disprove constructively) that there are no non-trivial deformation except Kaluza--Klein theory.


\section*{\sc Acknowledgments}
\vskip -2.7mm

The authors would like to thank Brando Bellazzini, Nicolas Boulanger, Chrysoula Markou, Paolo Pichini, Augusto Sagnotti and Evgeny Skvortsov  for useful discussions and suggestions.

\newpage

\begin{appendices}

\section{Notation and conventions} \label{app:nc}

We collect in this appendix the set of conventions used throughout the paper. Latin indices refer to 5D quantities, while Greek indices to 4D ones. The Minkowski metric is denoted with $\eta$ and is taken in the mostly-plus convention. The 4D Levi--Civita tensor is denoted with $\varepsilon^{\mu\nu\rho\sigma}$ and is such that $\varepsilon^{0123}=1$. Symmetrisation and antisymmetrisation of indices are denoted respectively with round and square brackets and are always intended with weight one. Moreover, we will employ the notation $T_{\mu]\dots[\nu}\equiv T_{[\mu|\dots|\nu]}$ when needed.

The 4D Clifford algebra is spanned by the Dirac matrices
\begin{align}
    \gamma^0=&i\sigma_1\otimes\mathbb{1},& \gamma^i=&\sigma_2\otimes\sigma^i, & \gamma_5=&\sigma_3\otimes\mathbb{1}\ .
\label{eqn:4D_Dirac_mat}
\end{align}

\noindent where $\sigma^i=\left\{\sigma_1,\sigma_2,\sigma_3\right\}$ are the  Pauli matrices. The associated charge conjugation matrix $C_4$ is taken to be
\begin{align}
	C_4=\sigma_3\otimes\sigma_2 \ , && C_4^{-1}=C_4^\textup{T}=C_4^\dagger=-C_4 \ , 
\end{align}

\noindent and it satisfies a symmetry condition
\begin{align}
    \(C_4\gamma^{\mu_1\dots \mu_r}\)^\textup{T}=&-t_rC_4\gamma^{\mu_1\dots \mu_r},
\label{eqn:4D_Cgamma_sym}
\end{align}

\noindent with $t_r=-1$ for $r=1,2\,[4]$ and $t_r=1$ for $r=0,3\,[4]$. The 4D Majorana fermions are then defined by the condition
\beq
    \bpsi=\psi^\textup{T}C_4.
\eeq

\noindent The $\gamma_5$ matrix in Eq.~\eqref{eqn:4D_Dirac_mat} allows to define the 4D chirality projectors and their action as
\begin{align}
    P_{\text{R},\text{L}}\equiv\frac{1\pm\gamma_5}{2}, && \psi_{\text{R},\text{L}}\equiv P_{\text{R},\text{L}}\psi, &&\bpsi_{\text{R},\text{L}}\equiv \bpsi P_{\text{R},\text{L}}=\overline{\psi_{\text{L},\text{R}}}\;.
\end{align}

In Section \ref{sec:stuckelberg} and in Appendix \ref{app:kk} we work with five-dimensional pure supergravity, in which the gravitino is actually a doublet of the $\mathrm{SU}(2)_R$ $R$-symmetry of the theory. These fermions are associated to the so called symplectic Clifford algebra, described by the following 5D symplectic Dirac matrices:
\begin{align}
    \gamma^0&=\mathbb{1}\otimes\(i\sigma_1\otimes\mathbb{1}\),\qquad\gamma^i=\mathbb{1}\otimes\(\sigma_2\otimes\sigma^i\),\qquad\gamma^4=\mathbb{1}\otimes\(\sigma_3\otimes\mathbb{1}\) \ ,
\label{eqn:5D_symp_gamma}
\end{align}

\noindent where the elements inside the parenthesis correspond to the usual Dirac indices while the elements in front correspond to the $\mathrm{SU}(2)_R$ indices. The charge conjugation matrix $\mathcal{C}_5$ is given in this setup by
\begin{equation}
    \mathcal{C}_5=\sigma_2\otimes\(\mathbb{1}\otimes\sigma_2\)\equiv \sigma_2\otimes \(\gamma_5C_4\) \ , 
\label{eqn:5D_C_mat}
\end{equation}

\noindent and it is such that
\beq
    \mathcal{C}_5^{-1}=\mathcal{C}_5^\textup{T}=\mathcal{C}_5^\dagger=\mathcal{C}_5.
\eeq

\noindent By construction, this symplectic Clifford algebra possess the symmetry property
\begin{align}
    \(\mathcal{C}_5\gamma^{M_1\dots M_r}\)^\textup{T}=&-t_r\mathcal{C}_5\gamma^{M_1\dots M_r},
\label{eqn:5D_Cgamma_sym}
\end{align}

\noindent with $t_r=-1$ for $r=0,1\,[4]$ and $t_r=1$ for $r=2,3\,[4]$. Thus, a symplectic-Majorana fermion $\lambda$ is defined by the condition
\begin{equation}
\bar\lambda=\lambda^\textup{T} \mathcal{C}_5.
\label{eqn:symp_maj}
\end{equation}

\noindent In Section \ref{sec:stuckelberg} and in Appendix \ref{app:kk} we will work directly with symplectic-Majorana spinors like \eqref{eqn:symp_maj} as long as five-dimensional quantities are concerned. We will then open the $\mathrm{SU}(2)_R$ components, which we denote with $\lambda^i$, with $i=1,2$, when needed.

\section{The Kaluza--Klein reduction of the 5D supergravity algebra on \texorpdfstring{${S^1/\mathbb{Z}_2}$}{S1/Z2}}\label{app:kk}

This appendix is devoted to the detailed Kaluza--Klein reduction of the 5D pure supergravity algebra on the orbifold $S^1/\mathbb{Z}_2$, from which it is possible to extract the $D=4$, $\N=1$ massive spin-2 supermultiplet of Eq.~\eqref{eqn:ms2_susy}. The field content of 5D pure supergravity is given by the f\"unfbein $E^A$, one abelian gauge field $\mathcal{A}$ and one  $\mathrm{SU}(2)_R$-Majorana gravitino $\Psi$. At the linearised level -- at which we work all along in this computation since it is enough to derive the desired multiplet -- the local supersymmetry transformations characterising this theory are \cite{Cremmer:1980gs,Chamseddine:1980mpx,Gunaydin:1983bi, Ceresole:2000jd}
\beq
\begin{aligned}
&\delta E^A=\frac{1}{2}\bar{\varepsilon}\gamma^A\Psi,  \\
&\delta \A = -\frac{i}{2}\bar{\varepsilon}\Psi , \\
&\delta \Psi= \mathcal{D}\varepsilon +i E^A\(\frac{1}{8}\gamma_{A}{}^{BC}\F_{BC}-\frac{1}{2}\F_{AB}\gamma^B\)\varepsilon,
\end{aligned}
\label{eqn:5D_susy}
\eeq

\noindent where $\F_{MN}=2\partial_{[M}\A_{N]}$ is the abelian gauge field strength, $\mathcal{D}_M$ is the five-dimensional Lorentz-covariant derivative and $\varepsilon$ is the 5D supersymmetry parameter. To perform the desired dimensional reduction, we start from the compactification ansatz
\beq
\begin{aligned}
&E^a=\phi^{-\frac{1}{2}}e^a, \\
&E^4=\phi\(dy+B\),  \\
&\A=a\(dy+B\)+A, \\
&\Psi=\phi^{\frac{5}{4}}\,\zeta\(dy+B\)+\phi^{-\frac{1}{4}}\(\psi-\frac{1}{2}e^a\gamma_a\gamma_5\zeta\),
\end{aligned}
\label{eqn:comp_ansatz}
\eeq

\noindent in which $y$ is the compact dimension and $e^a$ is the vierbein, $A$ and $B$ are 4D abelian gauge fields, of field strengths $F=dA$ and $G=dB$, $a$ and $\phi$ are two real scalar fields, and $\psi$ and $\zeta$ are two $\mathrm{SU}(2)_R$-Majorana fermions. Alongside this ansatz, it is convenient to redefine also the five dimensional supersymmetry parameter $\varepsilon$ as
\beq
\varepsilon=\phi^{-\frac{1}{4}}\epsilon \ ,
\label{eqn:4D_susy_param}
\eeq

\noindent with $\epsilon$ that will give the supersymmetry parameter in four dimensions.

The next step of this procedure is to plug the compactification ansatz \eqref{eqn:comp_ansatz} into the 5D algebra \eqref{eqn:5D_susy}, which yields its decomposition in terms of the 4D parameterization of \eqref{eqn:comp_ansatz}. In order to do this, we need the 4D decomposition along the ansatz \eqref{eqn:comp_ansatz} of the components of the 5D gauge field strength $\F_{MN}$ and spin-connection $\(\omega_M\)^A{}_B$. The former is given by
\begin{align}
    \F_{ab}&=\phi e_a{}^\mu e_b{}^\nu F_{\mu\nu}, & \F_{a4}&=\phi^{-\frac{1}{2}}e_a{}^\mu\(\partial_\mu a-\partial_yA_\mu\) .
\end{align}

\noindent The latter is more conveniently expressed through the anholonomy coefficients $\Omega_{AB}{}^C$, according to 
\begin{align}
  dE^A=\frac{1}{2}\Omega_{BC}{}^A{}E^B\wedge E^C,  && \(\omega_C\)_{AB}=\frac{1}{2}\(\Omega_{CAB}-\Omega_{CBA}+\Omega_{BAC}\),
\end{align}

\noindent which decompose as
\begin{equation}
    \begin{aligned}
        \Omega_{bc}{}^a&=\phi^{\frac{1}{2}}\Omega^{(4)}_{bc}{}^a-\phi^{-\frac{1}{2}}\partial_\mu\phi e_{[b}{}^\mu \delta^a_{c]}, &&&&& \Omega_{ab}{}^4&=\phi^2\,e_{[a}{}^\mu e_{b]}{}^\nu G_{\mu\nu}, \\
	   \Omega_{b4}{}^a&=\frac{1}{2}\phi^{-2}\partial_y\phi\,\delta^a_b-\phi^{-1}e_b{}^\mu\partial_y e_\mu{}^a, &&&&&	\Omega_{a4}{}^4&=\phi^{-\nicefrac{1}{2}}\,e_a{}^\mu\(\partial_\mu\phi-\phi\partial_y B_\mu\).
    \end{aligned}
\end{equation}

\noindent where $\Omega^{(4)}_{bc}$ are the anholonomy coefficients in 4D. Moreover, consistency between the compactification ansatz \eqref{eqn:comp_ansatz} and the structure of the original 5D supergravity algebra \eqref{eqn:5D_susy} requires, when performing the reduction, to implement the latter with an additional 5D local Lorentz transformation. At the linearised level, such transformations affects only the f\"unfbein, on which it acts as
\begin{equation}
    \delta_\textup{LL}E^A=\Lambda^A{}_B E^B,
\end{equation}

\noindent and the explicit transformation required is given by
\begin{align}
    \Lambda^a{}_b=\frac{1}{4}\phi^{\frac14}\(\bar{\varepsilon}\gamma^a{}_b\gamma_5\zeta\), && \Lambda^a{}_4 =\frac{1}{4}\phi^{\frac14}\(\bar{\varepsilon}\gamma^a\zeta\) \ .
\end{align}

The 4D supersymmetry transformations resulting from this procedure are then
\beq
\begin{aligned}
&\delta e_\mu{}^a=\frac{1}{2}\bar{\epsilon}\gamma^a\psi_\mu, \\
&\delta\phi=\frac{\phi}{2}\bar{\epsilon}\gamma_5\zeta, \\
&\delta a=-\frac{i}{2}\phi\bar{\epsilon}\zeta,  \\
&\delta A_\mu=-\frac{i}{2}\phi^{-\frac{1}{2}}\bar{\epsilon}\(\psi_\mu-\frac{1}{2}\gamma_\mu\gamma_5\zeta\), \\
&\delta B_\mu=\frac{\phi^{-\frac{3}{2}}}{2}\bar{\epsilon}\(\gamma_5\psi_\mu+\frac{3}{2}\gamma_\mu\zeta\), \\
&\begin{aligned}\delta\zeta=&-\frac{\phi^{-\frac{5}{2}}}{4}\partial_y\phi\epsilon+\frac{\phi^{-\frac{3}{2}}}{4}Q_{ab}\gamma^{ab}\epsilon+\frac{i}{8}\phi^{\frac{1}{2}}\(F_{\rho\sigma}+i\phi G_{\rho\sigma}\gamma_5\)\gamma^{\rho\sigma}\gamma_5\epsilon\\
&+\frac{i}{2}\phi^{-1}\[\partial_\rho a-\partial_yA_\rho+i\(\phi\partial_yB_\rho-\partial_\rho\phi\)\gamma_5\]\gamma^\rho \epsilon,
\end{aligned}\\
&\begin{aligned}\delta\psi_\mu=& D_\mu\epsilon-\frac{3}{8}\phi^{-\frac{5}{2}}\partial_y\phi\gamma_\mu\gamma_5\epsilon+\frac{\phi^{-\frac{3}{2}}}{2}e^a_{\mu}\(P_{ab}\gamma^b+\frac{1}{4}Q_{bc}\gamma_a\gamma^{bc}\)\gamma_5\epsilon \\
&-\frac{1}{4}\partial_yB_\rho\gamma_\mu\gamma^\rho\epsilon+\frac{3i}{4}\phi^{-1}\(\partial_yA_\mu-\partial_\mu a\)\gamma_5\epsilon-\frac{i}{8}\phi^{\frac{1}{2}}\(3F_{\mu\rho}^{+,5}-i\phi G_{\mu\rho}^{+,5}\gamma_5\)\gamma^\rho\epsilon.
\end{aligned}
\end{aligned}
\label{eqn:4d_5d_susy}
\eeq

\noindent In these formulae, $D_\mu$ is the 4D Lorentz-covariant derivative, while $\epsilon$ is the 4D supersymmetry parameter defined in \eqref{eqn:4D_susy_param}. The quantities $P_{ab}$ and $Q_{ab}$ are defined to be
\begin{align}
P_{ab}\equiv e_{(a}{}^\lambda\partial_y e_{b)}{}_{\lambda}, && Q_{ab}\equiv e_{[a}{}^\lambda\partial_y e_{b]}{}_{\lambda},
\end{align}

\noindent while $F_{\mu\nu}^{\pm,5}$ and $G_{\mu\nu}^{\pm,5}$ are the (anti)self-dual field strengths dressed by a $\gamma_5$, defined as
\begin{align}
F_{\mu\nu}^{\pm,5}\equiv F_{\mu\nu}\pm i\tilde{F}_{\mu\nu}\gamma_5 , && \tilde{F}^{\mu\nu}\equiv\frac{1}{2}\varepsilon^{\mu\nu\rho\sigma}F_{\rho\sigma}.
\label{eqn:dual_fs}
\end{align}

The transformations \eqref{eqn:4d_5d_susy} still carry the full dependence on the $5^\textup{th}$ compact coordinate $y$. Thus, the next step is to properly expand each field in modes along the $S^1/\mathbb{Z}_2$ interval spanned by $y$, which will further decompose the transformations \eqref{eqn:4d_5d_susy} into massless and massive transformation rules, properly organized in $\N=1$ multiplets. This mode expansion has to be consistent with the $\mathbb{Z}_2$ parity along $y$ modded out by the orbifold projection under consideration, namely with the transformation of each field under $\mathbb{Z}_2$, fixed by the invariance of the lagrangian.\footnote{For the explicit expression of the lagrangian of pure 5D supergravity, see \cite{Cremmer:1980gs,Chamseddine:1980mpx,Gunaydin:1983bi, Ceresole:2000jd}.} Starting from the bosons, the vierbein $e_\mu{}^a$ and the scalars $\phi$ and $a$ are even under $\mathbb{Z}_2$, while the gauge vectors $A_\mu$ and $B_\mu$ are odd: this fixes their Kaluza--Klein expansion to be\footnote{The expansion of the field $\phi$ is fixed to the one in \eqref{eqn:kk_bosons} by the additional requirement of being an exponential function.}
\begin{equation}
\begin{aligned}
e_\mu{}^a(y)&=e^{\scalebox{0.6}{$(0)$}}_\mu{}^a(x)+\sum_{n=1}^{\infty}e^{\scalebox{0.6}{$(n)$}}_\mu{}^a(x)\cos\(\frac{ny}{R}\), &  B_\mu(y)&=\sum_{n=1}^{\infty}B_\mu^{(n)}(x)\sin\(\frac{ny}{R}\),  \\
\phi(y)&=\phi^{(0)}(x)\[1+\sum_{n=1}^{\infty}\phi^{(n)}(x)\cos\(\frac{ny}{R}\)\], &
A_\mu(y)&=\sum_{n=1}^{\infty}A_\mu^{(n)}(x)\sin\(\frac{ny}{R}\), \\
a(y)&=a^{(0)}(x)+\sum_{n=1}^{\infty}a^{(n)}(x)\cos\(\frac{ny}{R}\). 
\end{aligned}
\label{eqn:kk_bosons}
\end{equation}

\noindent The projection on the fermions is instead more involved. The action of $\mathbb{Z}_2$ is in fact chiral and breaks therefore the original $\mathrm{SU}(2)_R$ symmetry:
\begin{align}
\Psi_i{}_\mu(-y)=\sigma^3_{ij}\gamma_5\Psi_j{}_\mu(y), && \Psi_i{}_5(-y)=-\sigma^3_{ij}\gamma_5\Psi_j{}_5(y),
\end{align}

\noindent so that
\beq
\begin{aligned}
\psi_1{}_\mu(-y)&=\gamma_5\psi_1{}_\mu(y), & \zeta_1{}_\mu(-y)&=-\gamma_5\zeta_1{}_\mu(y), \\
\psi_2{}_\mu(-y)&=-\gamma_5\psi_2{}_\mu(y), & \zeta_2{}_\mu(-y)&=\gamma_5\zeta_2{}_\mu(y).
\end{aligned}
\eeq

\noindent Then, the Kaluza--Klein mode expansion for the fermions result to be
\beq
\begin{aligned}
\psi_{1\mu}&=\psi_{1\mu \text{R}}+\psi_{1\mu \text{L}}=\psi^{(0)}_{1\mu \text{R}}+\sum_{n=1}^{\infty}\[\psi^{(n)}_{1\mu \text{R}}\cos\(\frac{ny}{R}\)+i\psi^{(n)}_{1\mu \text{L}}\sin\(\frac{ny}{R}\)\], \\
\psi_{2\mu}&=\psi_{1\mu \text{L}}+\psi_{1\mu \text{R}}=\psi^{(0)}_{2\mu \text{L}}+\sum_{n=1}^{\infty}\[\psi^{(n)}_{2\mu \text{L}}\cos\(\frac{ny}{R}\)+i\psi^{(n)}_{2\mu \text{R}}\sin\(\frac{ny}{R}\)\],\\
\zeta_1&=\zeta_{1\text{L}}+\zeta_{1\text{R}}=\zeta^{(0)}_{1\text{L}}+\sum_{n=1}^{\infty}\[\zeta^{(n)}_{1\text{L}}\cos\(\frac{ny}{R}\)+i\zeta^{(n)}_{1\text{R}}\sin\(\frac{ny}{R}\)\],  \\
\zeta_2&=\zeta_{2\text{R}}+\zeta_{2\text{L}}=\zeta^{(0)}_{2\text{R}}+\sum_{n=1}^{\infty}\[\zeta^{(n)}_{2\text{R}}\cos\(\frac{ny}{R}\)+i\zeta^{(n)}_{2\text{L}}\sin\(\frac{ny}{R}\)\]. 
\end{aligned}
\label{eqn:kk_fermions}
\eeq

\noindent This decomposition is consistent with the original symplectic Majorana condition, which at the level of the modes fixes, for every $n\ge 0$,
\begin{align}
    \zeta^{(n)}_{2\text{R}}=-\(\zeta^{(n)}_{1\text{L}}\)^*, && \zeta^{(n)}_{2\text{L}}=-\(\zeta^{(n)}_{1\text{R}}\)^*, &&\psi^{(n)}_{2\mu \text{R}}=\(\psi^{(n)}_{1\mu \text{L}}\)^*, && \psi^{(n)}_{2\mu \text{L}}=\(\psi^{(n)}_{1\mu \text{R}}\)^*.
\label{eqn:simp_maj}
\end{align}

Also the (local) supersymmetry parameter undergoes an analogous Kaluza--Klein decomposition and the associated zero mode correspond to the actual 4D supersymmetry parameters. The action of the $\mathbb{Z}_2$ symmetry is
\begin{equation}
    \epsilon_i(-y)=\sigma^3_{ij}\gamma_5\epsilon_j(y),
\label{eqn:susy_p_z2}
\end{equation}

\noindent where $\sigma^3$ is the third Pauli matrix. The actual 4D supersymmetry parameters correspond to the zero modes of $\epsilon_i(y)$ (the massive ones are discarded), which the $\mathbb{Z}_2$ parity \eqref{eqn:susy_p_z2} constrains to be
\begin{equation}
\begin{aligned}
\epsilon_1=\epsilon_{1\text{R}}, &&&&&& \epsilon_2=\epsilon_{2\text{L}},
\end{aligned}
\label{eqn:epsilon_chirality}
\end{equation}

\noindent with $\epsilon_{2\text{L}}=\(\epsilon_{1\text{R}}\)^*$, consistently with the five dimensional symplectic-Majorana condition \eqref{eqn:symp_maj}. This indicates therefore that the combination
\beq
\epsilon\equiv\epsilon_{1\text{R}}+\epsilon_{2\text{L}},
\label{eqn:4d_susy_p_def}
\eeq

\noindent that defines a 4D Majorana fermion, is the actual supersymmetry parameter in four dimension, showing explicitly how the orbifold projection breaks the original $\N=2$ supersymmetry in 5D to $\N=1$ in 4D.

Plugging the expansions \eqref{eqn:kk_bosons} and \eqref{eqn:kk_fermions} into \eqref{eqn:4d_5d_susy} yields the transformation rules associated to every Kaluza--Klein mode of the various fields, both massless and massive, which we now discuss separately.

\subsection{The massless sector}

The supersymmetry transformations of the massless modes result to be
\begin{equation}
    \begin{aligned}
        &\delta e^{\scalebox{0.6}{$(0)$}}_\mu{}^a=\frac{1}{2}\(\bar{\epsilon}_1\gamma^a\psi^{(0)}_{1\mu}+\bar{\epsilon}_2\gamma^a\psi^{(0)}_{2\mu}\), \\
        &\delta\phi^{(0)}=\frac{\phi^{(0)}}{2}\(\bepsilon_2\zeta^{(0)}_2-\bepsilon_1\zeta^{(0)}_1\), \\
        &\delta a^{(0)}=-\frac{i}{2}\phi^{(0)}\(\bepsilon_1\zeta^{(0)}_1+\bepsilon_2\zeta^{(0)}_2\), \\
        &\delta\psi^{(0)}_{1\mu R}=D_\mu\epsilon_1-\frac{3i}{4} {\phi^{\scalebox{0.6}{$(0)$}}}^{-1}  \partial_\mu a^{(0)} \epsilon_1,  \\
        &\delta\psi^{(0)}_{2\mu L}=D_\mu\epsilon_2+\frac{3i}{4}{\phi^{\scalebox{0.6}{$(0)$}}}^{-1}\partial_\mu a^{(0)} \epsilon_2,\\
        &\delta\zeta^{(0)}_{2R}=\frac{{\phi^{\scalebox{0.6}{$(0)$}}}^{-1}}{2}\partial_\rho\(\phi^{(0)}+ia^{(0)}\)\gamma^\rho\epsilon_2, \\
        &\delta\zeta^{(0)}_{1L}=-\frac{{\phi^{\scalebox{0.6}{$(0)$}}}^{-1}}{2}\partial_\rho\(\phi^{(0)}-ia^{(0)}\)\gamma^\rho\epsilon_1,
    \end{aligned}
\label{eqn:kk_m0_susy_1}
\end{equation}

\noindent and they correspond to pure $\N=1$ supergravity coupled to a chiral multiplet. This structure becomes manifest once the fields are redefined in a proper way. Equations \eqref{eqn:simp_maj} and \eqref{eqn:4d_susy_p_def} suggest to recombine the Majorana massless fermions as
\begin{align}
    \psi_{\mu}\equiv \psi^{(0)}_{1\mu R}+ \psi^{(0)}_{2\mu L}, && \zeta \equiv \phi^{(0)}\(\zeta^{(0)}_{2R}-\zeta^{(0)}_{1L}\).
\end{align}

\noindent These redefinitions, together with
\begin{equation}
    T\equiv \phi^{(0)}+ia^{(0)},
\end{equation}

\noindent and the renaming $e^{\scalebox{0.6}{$(0)$}}_{\mu}{}^a\to e_{\mu}{}^a$, fix the transformations of the massless sector to be
\beq
\begin{aligned}
&\delta e_{\mu}{}^a=\frac{1}{2}\bepsilon\gamma^a\psi_{\mu}, \\
&\delta\psi_{\mu}=D_\mu\epsilon+\frac{i}{2}Q_\mu\gamma_5\epsilon, \\
&\delta T=\bepsilon_\text{R}\zeta_\text{R},  \\
&\delta\zeta_\text{R}=\frac{1}{2}\cancel{\partial} T\epsilon_\text{L},
\end{aligned}
\eeq

\noindent which are the transformations of pure $D=4$, $\N=1$ supergravity $\left\{e_{\mu}{}^a, \psi_{\mu}\right\}$ coupled to chiral multiplet $\left\{T, \zeta \right\}$ \cite{Freedman:2012zz,DallAgata:2021uvl}, with K\"ahler potential and associated K\"ahler-covariantisation given by
\begin{align}
K=-3\log\(T+\overline{T}\), && Q_\mu=-\frac{3}{2i\(T+\overline{T}\)}\(\partial_\mu T-\partial_\mu\overline{T}\).
\end{align}

\noindent This is a no-scale K\"ahler potential expected in Kaluza--Klein reductions of this type.

\subsection{The massive sector}

We then move to the massive sector. We define the Kaluza--Klein mass of every mode as
\begin{align}
m_n={\phi^{\scalebox{0.6}{$(0)$}}}^{-\frac{3}{2}}\frac{n}{R},
\end{align}

\noindent with $R$ being the length of the compact dimension. Using then the formulae
\begin{equation}
\begin{aligned}
\partial_y\phi&=-{\phi^{\scalebox{0.6}{$(0)$}}}^{\frac52}\sum_nm_n\phi^{(n)}\sin\(\frac{ny}{R}\), &&& P_{ab}&=-{\phi^{\scalebox{0.6}{$(0)$}}}^{\frac32}\sum_n m_n e^{\scalebox{0.6}{$(0)$}}_{(a}{}^\rho e^{\scalebox{0.6}{$(n)$}}_{b)}{}_{\rho} \sin\(\frac{ny}{R}\),\\
\partial_yB_\mu&={\phi^{\scalebox{0.6}{$(0)$}}}^{\frac32}\sum_n m_nB^{(n)}_\mu\cos\(\frac{ny}{R}\),  &&& Q_{ab}&=-{\phi^{\scalebox{0.6}{$(0)$}}}^{\frac32}\sum_n m_n e^{\scalebox{0.6}{$(0)$}}_{[a}{}^\rho e^{\scalebox{0.6}{$(n)$}}_{b]}{}_{\rho} \sin\(\frac{ny}{R}\), \\
\partial_yA_\mu&={\phi^{\scalebox{0.6}{$(0)$}}}^{\frac32}\sum_n m_nA^{(n)}_\mu\cos\(\frac{ny}{R}\), &&& \omega^{\scalebox{0.6}{$(n)$}}_\mu{}^{ab}&=2e^{\scalebox{0.6}{$(0)$}}{}^{[a|}{}^{\alpha}\partial^{\scalebox{0.6}{\color{white}$(n)$}}_{[\mu}e^{\scalebox{0.6}{$(n)$}}_{\alpha]}{}^{|b]}-e^{\scalebox{0.6}{$(0)$}}_\alpha{}^{[a}e^{\scalebox{0.6}{$(0)$}}_\beta{}^{ b]}e^{\scalebox{0.6}{$(0)$}}_\mu{}^c\partial^\alpha e^{\scalebox{0.6}{$(n)$}}_c{}^{\beta}, 
\end{aligned}
\end{equation}

\noindent one obtains that the massive modes of the set of supersymmetry transformations \eqref{eqn:4d_5d_susy}. The bosonic transformation rules are  
\beq
\begin{aligned}
&\delta e_n{}^a_\mu=\frac{1}{2}\(\bar{\epsilon}_1\gamma^a\psi^{(n)}_{1\mu}+\bar{\epsilon}_2\gamma^a\psi^{(n)}_{2\mu }\), \\
&\delta\phi^{(n)}=\frac{1}{2}\(\bepsilon_2\zeta^{(n)}_2-\bepsilon_1\zeta^{(n)}_1\), \\
&\delta a^{(n)}=-\frac{i}{2}\phi^{(0)}\(\bepsilon_1\zeta^{(n)}_1+\bepsilon_2\zeta^{(n)}_2\),  \\
&\delta A^{(n)}_\mu=\frac{{\phi^{\scalebox{0.6}{$(0)$}}}^{-\frac12}}{2}\[\bepsilon_1\(\psi^{(n)}_{1\mu}-\frac{1}{2}\gamma_\mu\zeta^{(n)}_1\)+\bepsilon_2\(\psi^{(n)}_{2\mu}-\frac{1}{2}\gamma_\mu\zeta^{(n)}_2\)\], \\
&\delta B^{(n)}_\mu=\frac{i}{2}{\phi^{\scalebox{0.6}{$(0)$}}}^{-\frac{3}{2}}\[\bepsilon_2\(\psi^{(n)}_{2\mu}+\frac{3}{2}\gamma_\mu\zeta^{(n)}_2\)-\bepsilon_1\(\psi^{(n)}_{1\mu}-\frac{3}{2}\gamma_\mu\zeta^{(n)}_1\)\],
\end{aligned}
\label{eqn:susy_bare_m_B}
\end{equation}

\noindent while the transformations on the massive spin-$\textstyle{\frac12}$ and spin-$\textstyle{\frac32}$ are respectively
\beq
\begin{aligned}
&\delta\zeta^{(n)}_{1\text{L}}=\frac{m_n}{2}\[-i\({\phi^{\scalebox{0.6}{$(0)$}}}^{\frac12}A^{(n)}_\rho-\frac{{\phi^{\scalebox{0.6}{$(0)$}}}^{-1}}{m_n}\partial_\rho a^{(n)}\)+\({\phi^{\scalebox{0.6}{$(0)$}}}^{\frac32}B^{(n)}_\rho-\frac{1}{m_n}\partial_\rho \phi^{(n)}\)\]\gamma^\rho\epsilon_1, \\
&\delta\zeta^{(n)}_{1\text{R}}=\frac{{\phi^{\scalebox{0.6}{$(0)$}}}^{\frac12}}{8}\(F^{(n)}_{\rho\sigma}+i\phi^{(0)}G^{(n)}_{\rho\sigma}\)\gamma^{\rho\sigma}\epsilon_1+\frac{i}{4}m_n \(e^{\scalebox{0.6}{$(0)$}}_a{}^\rho e^{\scalebox{0.6}{$(n)$}}_b{}_{\rho}\gamma^{ab}-\phi^{(n)}\)\epsilon_1, \\
&\delta\zeta^{(n)}_{2\text{L}}=-\frac{{\phi^{\scalebox{0.6}{$(0)$}}}^{\frac12}}{8}\(F^{(n)}_{\rho\sigma}-i\phi^{(0)}G^{(n)}_{\rho\sigma}\)\gamma^{\rho\sigma}\epsilon_2+\frac{i}{4}m_n\(e^{\scalebox{0.6}{$(0)$}}_a{}^\rho e^{\scalebox{0.6}{$(n)$}}_b{}_{\rho}\gamma^{ab}-\phi^{(n)}\)\epsilon_2, \\
&\delta\zeta^{(n)}_{2\text{R}}=\frac{m_n}{2}\[-i\({\phi^{\scalebox{0.6}{$(0)$}}}^{\frac12}A^{(n)}_\rho-\frac{{\phi^{\scalebox{0.6}{$(0)$}}}^{-1}}{m_n}\partial_\rho a^{(n)}\)-\({\phi^{\scalebox{0.6}{$(0)$}}}^{\frac32}B^{(n)}_\rho-\frac{1}{m_n}\partial_\rho \phi^{(n)}\)\]\gamma^\rho\epsilon_2,
\end{aligned}
\label{eqn:susy_bare_m_12}
\eeq

\noindent and 
\beq
\begin{aligned}
&\begin{aligned}
\delta\psi^{(n)}_{1\mu \text{R}}&=\frac{1}{4}\(2e^{\scalebox{0.6}{$(0)$}}{}^{a}{}^{\alpha}\partial^{\scalebox{0.6}{\color{white}$(n)$}}_{[\mu}e^{\scalebox{0.6}{$(n)$}}_{\alpha]}{}^{b}-e^{\scalebox{0.6}{$(0)$}}_\alpha{}^{a}e^{\scalebox{0.6}{$(0)$}}_\beta{}^{ b}e^{\scalebox{0.6}{$(0)$}}_\mu{}^c\partial^\alpha e^{\scalebox{0.6}{$(n)$}}_c{}^{\beta}\)\gamma_{ab}\epsilon_1-\frac{m_n}{4}{\phi^{\scalebox{0.6}{$(0)$}}}^{\frac32}B^{(n)}_\rho\gamma_\mu\gamma^\rho\epsilon_1\\
&+\frac{3i}{4}m_n\({\phi^{\scalebox{0.6}{$(0)$}}}^{\frac12}A^{(n)}_\mu-\frac{{\phi^{\scalebox{0.6}{$(0)$}}}^{-1}}{m_n}\partial_\mu a^{(n)}\)\epsilon_1,
\end{aligned}  \\
&\begin{aligned}
\delta\psi^{(n)}_{1\mu \text{L}}&=-\frac{{\phi^{\scalebox{0.6}{$(0)$}}}^{\frac12}}{8}\(3F^{(n)+,5}_{\mu\rho}+i\phi^{(0)}G^{(n)+,5}_{\mu\rho}\)\gamma^\rho\epsilon_1+\frac{i}{4}m_ng^{(n)}_{\mu\rho}\gamma^\rho\epsilon_1\\
&+\frac{i}{4}m_n\(e^{\scalebox{0.6}{$(0)$}}_a{}^\rho e^{\scalebox{0.6}{$(n)$}}_b{}_{\rho}\gamma_\mu\gamma^{ab}-\phi^{(n)}\gamma_\mu\)\epsilon_1,
\end{aligned} \\
&\begin{aligned}
\delta\psi^{(n)}_{2\mu \text{R}}&=-\frac{{\phi^{\scalebox{0.6}{$(0)$}}}^{\frac12}}{8}\(3F^{(n)+,5}_{\mu\rho}-i\phi^{(0)}G^{(n)+,5}_{\mu\rho}\)\gamma^\rho\epsilon_2-\frac{i}{4}m_ng^{(n)}_{\mu\rho}\gamma^\rho\epsilon_2\\
&-\frac{i}{4}m_n\(e^{\scalebox{0.6}{$(0)$}}_a{}^\rho e^{\scalebox{0.6}{$(n)$}}_b{}_{\rho}\gamma_\mu\gamma^{ab}-\phi^{(n)}\gamma_\mu\)\epsilon_2,
\end{aligned}  \\
&\begin{aligned}
\delta\psi^{(n)}_{2\mu \text{L}}&=\frac{1}{4}\(2e^{\scalebox{0.6}{$(0)$}}{}^{a}{}^{\alpha}\partial^{\scalebox{0.6}{\color{white}$(n)$}}_{[\mu}e^{\scalebox{0.6}{$(n)$}}_{\alpha]}{}^{b}-e^{\scalebox{0.6}{$(0)$}}_\alpha{}^{a}e^{\scalebox{0.6}{$(0)$}}_\beta{}^{ b}e^{\scalebox{0.6}{$(0)$}}_\mu{}^c\partial^\alpha e^{\scalebox{0.6}{$(n)$}}_c{}^{\beta}\)\gamma_{ab}\epsilon_2-\frac{m_n}{4}{\phi^{\scalebox{0.6}{$(0)$}}}^{\frac32}B^{(n)}_\rho\gamma_\mu\gamma^\rho\epsilon_2\\
&-\frac{3i}{4}m_n\({\phi^{\scalebox{0.6}{$(0)$}}}^{\frac12}A^{(n)}_\mu-\frac{{\phi^{\scalebox{0.6}{$(0)$}}}^{-1}}{m_n}\partial_\mu a^{(n)}\)\epsilon_2.
\end{aligned}
\end{aligned}
\label{eqn:susy_bare_m_32}
\eeq

The whole set massive degrees of freedom recombines, for each mode $n$, to a massive spin-1, two massive Majorana spin-$\textstyle{\frac32}$ and a massive spin-2 fields, according to the St\"uckelberg parameterisation, as customary for massive gauge fields. The St\"uckelberg symmetry comes naturally in dimensional compactifications, as the massive modes of the higher-dimensional gauge transformations.\footnote{See \cite{Hinterbichler:2011tt} for an explicit analysis of this property in the case of the massive spin-2 field.} To determine the combinations of the massive degrees of freedom of Eq.~\eqref{eqn:kk_bosons} and \eqref{eqn:kk_fermions} associated to the unitary gauge of the correspondent St\"uckelberg symmetry, we start from the simplest (and more standard) of such combinations, the one for the massive spin-1 field, which we take to be
\beq
    {A_\mu^{\scalebox{0.6}{$(n)$}}}{}^\textup{S}={\phi^{\scalebox{0.6}{$(0)$}}}^{\frac12}A^{(n)}_\mu-\frac{{\phi^{\scalebox{0.6}{$(0)$}}}^{-1}}{m_n}\partial_\mu a^{(n)},
\eeq

\noindent as suggested also by the massive spin-$\textstyle{\frac32}$ supersymmetry transformations \eqref{eqn:susy_bare_m_32}. Acting with a supersymmetry transformation \eqref{eqn:susy_bare_m_B}, one finds
\begin{equation}
    \delta  {A_\mu^{\scalebox{0.6}{$(n)$}}}{}^\textup{S}=\frac{1}{2}\[\bepsilon_1\(\psi^{(n)}_{1\mu \text{L}}-\frac{1}{2}\gamma_\mu\zeta^{(n)}_{1\text{R}}+\frac{i}{m_n}\partial_\mu\zeta^{(n)}_{1\text{L}}\)+\bepsilon_2\(\psi^{(n)}_{2\mu \text{R}}+\frac{1}{2}\gamma_\mu\zeta^{(n)}_{2\text{L}}+\frac{i}{m_n}\partial_\mu\zeta^{(n)}_{2\text{R}}\)\],
\end{equation}

\noindent which indicates that the correct St\"uckelberg combinations for $\psi^{(n)}_{1\mu \text{L}}$ and $\psi^{(n)}_{2\mu \text{R}}$ are
\begin{equation}
    \begin{aligned}
        {{\psi_{1\mu\text{L}}^{\scalebox{0.6}{$(n)$}}}}{}^\textup{S}&=\psi^{(n)}_{1\mu \text{L}}-\frac{1}{2}\gamma_\mu\zeta^{(n)}_{1\text{R}}+\frac{i}{m_n}\partial_\mu\zeta^{(n)}_{1\text{L}},\\ 
         {{\psi_{2\mu\text{R}}^{\scalebox{0.6}{$(n)$}}}}{}^\textup{S}&=\psi^{(n)}_{2\mu \text{R}}+\frac{1}{2}\gamma_\mu\zeta^{(n)}_{2\text{L}}+\frac{i}{m_n}\partial_\mu\zeta^{(n)}_{2\text{R}},
    \end{aligned}
\label{eqn:psi_1L_S}
\end{equation}

\noindent where $ {{\psi_{2\mu\text{R}}^{\scalebox{0.6}{$(n)$}}}}{}^\textup{S}=\( {{\psi_{1\mu\text{L}}^{\scalebox{0.6}{$(n)$}}}}{}^\textup{S}\)^*$ consistently, so that they can be combined in a Majorana spin-$\textstyle{\frac32}$ field
\beq
\begin{aligned}
    {\lambda_\mu^{\scalebox{0.6}{$(n)$}}}{}^\textup{S}\equiv& {{\psi_{1\mu\text{L}}^{\scalebox{0.6}{$(n)$}}}}{}^\textup{S}+ {{\psi_{2\mu\text{R}}^{\scalebox{0.6}{$(n)$}}}}{}^\textup{S}=\(\psi^{(n)}_{1\mu \text{L}}+\psi^{(n)}_{2\mu \text{R}}\)-\frac{1}{2}\gamma_\mu\(\zeta^{(n)}_{1\text{R}}-\zeta^{(n)}_{2\text{L}}\)+\frac{i}{m_n}\partial_\mu\(\zeta^{(n)}_{1\text{L}} +\zeta^{(n)}_{2\text{R}}\).
\end{aligned}
\eeq

\noindent Proceeding in this way, one finds the correct St\"uckelberg combinations for the remaining spin-$\textstyle{\frac32}$ and spin-2 fields ${\chi_\mu^{\scalebox{0.6}{$(n)$}}}{}^\textup{S}$ and ${g_{\mu\nu}^{\scalebox{0.6}{$(n)$}}}{}^\textup{S}$, which are 
\begin{align}
&{\chi_\mu^{\scalebox{0.6}{$(n)$}}}{}^\textup{S}=-i\left[\(\psi^{(n)}_{1\mu \text{R}}-\psi^{(n)}_{2\mu \text{L}}\)-\frac{1}{2}\gamma_\mu\(\zeta^{(n)}_{1\text{L}}+\zeta^{(n)}_{2\text{R}}\)+\frac{i}{m_n}\partial_\mu\(\zeta^{(n)}_{1\text{R}}-\zeta^{(n)}_{2\text{L}}\)\right], \\
&{g_{\mu\nu}^{\scalebox{0.6}{$(n)$}}}{}^\textup{S}=2e^{\scalebox{0.6}{$(0)$}}_{(\mu}{}^a e^{\scalebox{0.6}{$(n)$}}_{\nu)}{}_a-\phi^{(n)}g^{(0)}_{\mu\nu}+\frac{{\phi^{\scalebox{0.6}{$(0)$}}}^{\frac32}}{m_n}\(\partial_\mu B^{(n)}_\nu+\partial_\nu B^{(n)}_\mu\)-\frac{2}{m_n^2}\partial_\mu\partial_\nu\phi^{(n)}.
\end{align}

\noindent which together with  ${A_\mu^{\scalebox{0.6}{$(n)$}}}{}^\textup{S}$ and $ {\lambda_\mu^{\scalebox{0.6}{$(n)$}}}{}^\textup{S}$ form the massive spin-2 supermultiplet we were looking for. The linearised supersymmetry transformations of the above combinations, where one retains only the coupling to the massless modes of the vierbein $e_\mu{}^a$ and the scalar $\phi$ (but not their derivatives) and discards all other couplings, give us the algebra of a rigid massive spin-2 supermultiplet of mass $m_n$, for each $n$ mode of the Kaluza--Klein tower. These are the transformations given in Eq.~\eqref{eqn:ms2_susy}, which can be thought of independently from the Kaluza--Klein reduction used to obtain them.

\section{The improvement to the axial current}\label{app:Dj}

In this appendix we describe how the most general improvement to the axial current of Eq.~\eqref{eqn:Dj_fin} is fixed. 
%
The first requirement this ansatz has to satisfy is to have the correct spacetime parity of the axial current, which is a pseudo-vector: 
\beq
j^\mu\to -\mathcal{P}^\mu{}_\nu j^\nu.
\eeq


\noindent Consistency with such parity constraints $L^{\mu\nu}$ to include the following operators: 
\beq
\begin{aligned}
    L^{\mu\nu}=&c_\textup{B1} h^{[\mu}{}_\lambda \partial^\lambda A^{\nu]}+c_\textup{B2}h^{[\mu}{}_\lambda\partial^{\nu]}A^\lambda+\frac{c_\textup{B3}}{m}\varepsilon^{\mu\nu\rho\sigma}\partial_\lambda A_\rho\partial_\sigma A^\lambda\\
    &+c_\textup{F1}\varepsilon^{\mu\nu\rho\sigma}\bchi_\rho\psi_\sigma+ic_\textup{F2}\bchi^{[\mu}\gamma_5\psi^{\nu]}.
\end{aligned}
\label{eqn:Dj_1}
\eeq

\noindent Then, such an improvement has to satisfy the linear multiplet condition \cite{Ferrara:1974pz, Komargodski:2010rb} of Eq.~(\ref{eqn:linear_m_comp})
\begin{equation}
\delta^2_\textup{L}L^{\mu\nu}=0,
\end{equation}

\noindent as stated in Eq.~(\ref{eqn:Dj_lin_m}). Using the Fierz relation
\begin{equation}
\epsilon_\textup{L}\bepsilon_\textup{L}=-\frac{1}{2}\left(\bepsilon_\textup{L}\epsilon_\textup{L}\right)P_\textup{L},
\end{equation}

\noindent the transformation of the massive fields are
\begin{equation}
\begin{aligned}
\delta^2_\textup{L}A_\mu=&0, &&&&&&& \delta^2_\textup{L}\psi_\mu=&-\frac{m}{2}\(\bepsilon_\textup{L}\epsilon_\textup{L}\)\chi_{\mu\textup{R}}, \\
\delta^2_\textup{L}h_{\mu\nu}=&0, &&&&&&& \delta^2_\textup{L}\chi_\mu=&-\frac{m}{2}\(\bepsilon_\textup{L}\epsilon_\textup{L}\)\psi_{\mu\textup{L}},
\end{aligned}
\end{equation}

\noindent so that the various operators of $L^{\mu\nu}$ in (\ref{eqn:Dj_1}) transform as
\beq
\begin{aligned}
    \bullet& & &\delta_\textup{L}^2\left(h^{[\mu}{}_\lambda \partial^\lambda A^{\nu]}\)=\frac{i}{4}\(\bepsilon_\textup{L}\epsilon_\textup{L}\)\[-\frac{1}{m}\partial_\lambda\bpsi^{[\mu}_\textup{L}\partial^{\nu]}\psi^\lambda_{\textup{L}}+\bchi^\lambda_\textup{R}\gamma^{[\mu}\partial^{\nu]}\psi_{\lambda\textup{L}}\], \\
    \bullet& & &\delta_\textup{L}^2\left(h^{[\mu}{}_\lambda\partial^{\nu]}A^\lambda\right)=\frac{i}{4}\(\bepsilon_\textup{L}\epsilon_\textup{L}\)\[\frac{1}{m}\partial_\lambda\bpsi^{[\mu}_\textup{L}\partial^{\nu]}\psi^\lambda_{\textup{L}}+\bchi^\lambda_\textup{R}\gamma^{[\mu}\partial^{\nu]}\psi_{\lambda\textup{L}}\], \\
    \bullet& & &\delta_\textup{L}^2\left(\frac{1}{m}\varepsilon^{\mu\nu\rho\sigma}\partial_\lambda A_\rho\partial_\sigma A^\lambda\right)=\frac{i}{4}\(\bepsilon_\textup{L}\epsilon_\textup{L}\)\[\frac{2}{m}\partial_\lambda\bpsi^{[\mu}_\textup{L}\partial^{\nu]}\psi^\lambda_{\textup{L}}-2\bchi^\lambda_\textup{R}\gamma^{[\mu}\partial^{\nu]}\psi_{\lambda\textup{L}}\], \\
    \bullet& & &\delta_\textup{L}^2\(\varepsilon^{\mu\nu\rho\sigma}\bchi_\rho\psi_\sigma\)=0, \\
\bullet& & &\delta_\textup{L}^2\(\bchi^{[\mu}\gamma_5\psi^{\nu]}\)=0.
\end{aligned}
\label{eqn:d2L_Dj}
\eeq

Thus, the fermionic improvements are consistent on their own, while the bosonic ones have to appear in the combination defined by
\begin{equation}
\left\{\begin{aligned}
&c_\textup{B1}-c_\textup{B2}-2c_\textup{B3}=0\\
&c_\textup{B1}+c_\textup{B2}-2c_\textup{B3}=0
\end{aligned}\right. \quad \iff \quad c_\textup{B2}=0 \quad \text{\&} \quad c_\textup{B1}=2c_\textup{B3}\equiv 2c_\textup{B}.
\end{equation}

\noindent Therefore, the ansatz for the general improvement to the axial current is fixed to be
\begin{equation}
L^{\mu\nu}=c_\textup{B}\(2h^{[\mu|}{}_\lambda\partial^\lambda A^{|\nu]}+\frac{1}{m}\varepsilon^{\mu\nu\rho\sigma}\partial_\lambda A_\rho\partial_\sigma A^\lambda\)+c_\textup{F1}\varepsilon^{\mu\nu\rho\sigma}\bchi_\rho\psi_\sigma+ic_\textup{F2}\bchi^{[\mu}\gamma_5\psi^{\nu]},
\end{equation}

\noindent as in Eq.~(\ref{eqn:Dj_fin}).

\end{appendices}

\newpage
\bibliographystyle{JHEP}
\bibliography{refs}

\end{document}